\pdfoutput=1

\documentclass[11pt,twoside,a4paper,cmspaper,final,collab]{cms-tdr}
\def\svnVersion{422370P}\def\cmsCernNoTag{CERN-EP-2017-050}\def\cmsCernDate{\today}\def\cmsMessage{Published in Physical Review D as \href{http://dx.doi.org/10.1103/PhysRevD.96.032002}{\doi{10.1103/PhysRevD.96.032002.}}}

\begin{document}\cmsNoteHeader{TOP-15-008}

\hyphenation{had-ron-i-za-tion}
\hyphenation{cal-or-i-me-ter}
\hyphenation{de-vices}
\RCS$Revision: 416172 $
\RCS$HeadURL: svn+ssh://svn.cern.ch/reps/tdr2/papers/TOP-15-008/trunk/TOP-15-008.tex $
\RCS$Id: TOP-15-008.tex 416172 2017-07-13 19:19:54Z nmirman $
\newlength\cmsFigWidth
\ifthenelse{\boolean{cms@external}}{\setlength\cmsFigWidth{0.987\columnwidth}}{\setlength\cmsFigWidth{0.66\textwidth}}
\newlength\cmsFigMultiWidth
\ifthenelse{\boolean{cms@external}}{\setlength\cmsFigMultiWidth{0.42\textwidth}}{\setlength\cmsFigMultiWidth{0.48\textwidth}}
\ifthenelse{\boolean{cms@external}}{\providecommand{\cmsLeft}{upper\xspace}}{\providecommand{\cmsLeft}{left\xspace}}
\ifthenelse{\boolean{cms@external}}{\providecommand{\cmsRight}{lower\xspace}}{\providecommand{\cmsRight}{right\xspace}}
\ifthenelse{\boolean{cms@external}}{\providecommand{\cmsLLeft}{Upper\xspace}}{\providecommand{\cmsLLeft}{Left\xspace}}
\ifthenelse{\boolean{cms@external}}{\providecommand{\cmsRRight}{Lower\xspace}}{\providecommand{\cmsRRight}{Right\xspace}}
\ifthenelse{\boolean{cms@external}}{\providecommand{\NA}{\ensuremath{\cdots}\xspace}}{\providecommand{\NA}{\ensuremath{\text{---}}\xspace}}
\ifthenelse{\boolean{cms@italic}}{\newcommand{\cmsBoldSymbolFace}{\boldsymbol}}{\newcommand{\cmsBoldSymbolFace}{\mathbf}}

\ifthenelse{\boolean{cms@external}}{\renewcommand{\mtt}{\ensuremath{M_{\mathrm{T}2}}}}{\newcommand{\mtt}{\ensuremath{M_{\mathrm{T}2}}}}
\newcommand{\mtr}{\ensuremath{M_{\mathrm{T}}}}
\newcommand{\mbl}{\ensuremath{M_{\PQb\ell}}}
\newcommand{\mblv}{\ensuremath{M_{\PQb\ell\nu}}}
\newcommand{\mt}{\ensuremath{M_{\PQt}}\xspace}
\newcommand{\mtmc}{\ensuremath{M_{\PQt}^{\text{MC}}}}
\newcommand{\mw}{\ensuremath{M_{\PW}}}
\newcommand{\mnu}{\ensuremath{M_{\nu}}}
\newcommand{\mttbb}{\ensuremath{M_{\mathrm{T}2}^{\PQb\PQb}}}
\newcommand{\mttbl}{\ensuremath{M_{\mathrm{T}2}^{\PQb\ell}}}
\newcommand{\mttll}{\ensuremath{M_{\mathrm{T}2}^{\ell\ell}}}
\newcommand{\mttbbb}{\ensuremath{\boldsymbol{M}_{\text{\textbf{T2}}}^{\cmsBoldSymbolFace{bb}}}}
\newcommand{\mttblb}{\ensuremath{\boldsymbol{M}_{\text{\textbf{T2}}}^{\cmsBoldSymbolFace{b}\boldsymbol{\ell}}}}
\newcommand{\mttllb}{\ensuremath{\boldsymbol{M}_{\text{\textbf{T2}}}^{\boldsymbol{\ell\ell}}}}
\newcommand{\gp}{\ensuremath{\mathbf{u}}}
\newcommand{\gptrain}{\ensuremath{\gp_{\text{train}}}}
\newcommand{\gptest}{\ensuremath{\gp_{\text{test}}}}
\newcommand{\jsf}{\ensuremath{\mathrm{JSF}}}
\newcommand{\bl}{\ensuremath{\PQb\ell}}
\newcommand{\blv}{\ensuremath{\PQb\ell\nu}}
\newcommand{\Irel}{\ensuremath{I_{\text{rel}}}}
\newcommand{\ptupst}{\ensuremath{\ptvec^{\text{ upst}}}}
\newcommand{\pta}{\ensuremath{\ptvec^{\,\mathrm{a}}}}
\newcommand{\ptb}{\ensuremath{\ptvec^{\,\mathrm{b}}}}
\newcommand{\ptt}{\ensuremath{\ptvec^{\,2}}}
\newcommand{\Lt}{\ensuremath{\mathcal{L}}}
\newcommand{\sm}{\ensuremath{\sigma_m}}
\newcommand{\fa}{\ensuremath{f}\xspace}
\newcommand{\PE}{pseudo-experiment\xspace}
\newcommand{\PEs}{pseudo-experiments\xspace}
\newcommand{\ca}{\ensuremath{\circledast}}
\newcommand{\whyb}{\ensuremath{w_{\text{hyb}}}}
\newcommand{\mthyb}{\ensuremath{\mt^{\text{hyb}}}}
\newcommand{\mtD}{\ensuremath{\mt^{\text{1D}}}}
\newcommand{\mtDD}{\ensuremath{\mt^{\text{2D}}}}
\newcommand{\mtMAOS}{\ensuremath{\mt^{\text{MAOS}}}}
\newcommand{\jsfDD}{\ensuremath{\jsf^{\text{2D}}}}
\newcommand{\nevts}{\ensuremath{41\,640}\xspace}
\newcommand{\luminosity}{\ensuremath{19.7 \pm 0.5\fbinv}}
\newcommand{\fitDmt}{\ensuremath{172.39 \pm 0.17\stat \,^{+0.91}_{-0.95}\syst\GeV}}
\newcommand{\fitDDmt}{\ensuremath{171.56 \pm 0.46\stat \,^{+1.31}_{-1.25}\syst\GeV}}
\newcommand{\fitDDmtshort}{\ensuremath{171.56 \pm 0.46\stat \,^{+1.31}_{-1.25}\syst}}
\newcommand{\fitDDjsf}{\ensuremath{1.011 \pm 0.006\stat \,^{+0.015}_{-0.014}\syst}}
\newcommand{\fithybmt}{\ensuremath{172.22 \pm 0.18\stat \,^{+0.89}_{-0.93}\syst\GeV}}
\newcommand{\fitMAOSmt}{\ensuremath{171.54 \pm 0.19\stat \,^{+1.27}_{-1.02}\syst\GeV}}

\cmsNoteHeader{TOP-15-008}
\title{Measurement of the top quark mass in the dileptonic \texorpdfstring{\ttbar}{ttbar} decay channel using the mass observables \texorpdfstring{\mbl, \mtt, and \mblv}{Mbl, MT2, and Mblv} in pp collisions at \texorpdfstring{$\sqrt{s} = 8$\TeV}{sqrt(s) = 8 TeV}}

\date{\today}

\abstract{
A measurement of the top quark mass ($M_{\PQt}$) in the dileptonic $\ttbar$ decay channel is performed using data from proton-proton collisions at a center-of-mass energy of 8\TeV.  The data was recorded by the CMS experiment at the LHC and corresponds to an integrated luminosity of $19.7 \pm 0.5\fbinv$.  Events are selected with two oppositely charged leptons ($\ell=\Pe,\mu$) and two jets identified as originating from \PQb quarks.  The analysis is based on three kinematic observables whose distributions are sensitive to the value of $M_{\PQt}$.  An invariant mass observable, $M_{\PQb\ell}$, and a `stransverse mass' observable, $M_{\mathrm{T}2}$, are employed in a simultaneous fit to determine the value of $M_{\PQt}$ and an overall jet energy scale factor (JSF).  A complementary approach is used to construct an invariant mass observable, $M_{\PQb\ell\nu}$, that is combined with $M_{\mathrm{T}2}$ to measure $M_{\PQt}$.  The shapes of the observables, along with their evolutions in $M_{\PQt}$ and JSF, are modeled by a nonparametric Gaussian process regression technique.  The sensitivity of the observables to the value of $M_{\PQt}$ is investigated using a Fisher information density method.  The top quark mass is measured to be $172.22 \pm 0.18\stat\,^{+0.89}_{-0.93}\syst\GeV$.}

\hypersetup{%
pdfauthor={CMS Collaboration},%
pdftitle={Measurement of the top quark mass in the dileptonic ttbar decay channel using the mass observables Mbl, MT2, and Mblv  in pp collisions at sqrt(s) = 8 TeV},%
pdfsubject={CMS},%
pdfkeywords={CMS, physics, top, mass, MT2, MAOS, jet energy scale, Gaussian process, Fisher information}}

\maketitle

\section{Introduction}
\label{section:intro}

The top quark mass is a fundamental parameter of the standard model (SM), and an important component in global electroweak fits evaluating the self-consistency of the SM~\cite{gfitter}.
In addition, the value of \mt  has implications for the stability of the SM electroweak vacuum due to the role of the top quark in the quartic term of the Higgs potential \cite{vacuum_stability}.
Measurements of \mt  have been conducted by the CDF and \DZERO experiments at the Tevatron, and by the ATLAS and CMS experiments at the CERN LHC.
These measurements are typically calibrated against the top quark mass parameter in Monte Carlo (MC) simulation.  Studies suggest that this parameter can be related to the top quark mass in a theoretically well-defined scheme with a precision of about 1\GeV \cite{mtop_mc}.
A combination of measurements including all four experiments and \ttbar\ decay channels with zero, one, or two high-\pt electrons or muons (all-hadronic, semileptonic, and dileptonic, respectively) gives a value of $173.34 \pm 0.36\stat \pm 0.67\syst\GeV$ \cite{world_comb} for the top quark mass.
Currently, the most precise experimental determination of \mt  is provided by CMS using a combination of measurements in all \ttbar\ decay channels, yielding a value of $172.44 \pm 0.13\stat \pm 0.47\syst\GeV$~\cite{cms_comb}.
In the dileptonic \ttbar\ decay channel, the ATLAS \cite{atlas_dilepton} and CMS \cite{cms_comb} Collaborations have recently determined \mt  to be $172.99 \pm 0.41\stat \pm 0.74\syst\GeV$ and $172.82 \pm 0.19\stat \pm 1.22\syst\GeV$, respectively.
This paper presents a reanalysis of the dileptonic \ttbar data set recorded in 2012, with a primary motivation of reducing the systematic uncertainties in \mt  determination.

The dileptonic top quark pair (\ttbar) decay topology, $\ttbar\to (\PQb\ell^+\nu)(\PAQb\ell^-\overline{\nu})$, with $\ell=(\Pe,\mu)$, presents a challenge in mass measurement arising primarily from the presence of two neutrinos in the final state.  While the undetected \ptvec\ of a single final-state neutrino in a semileptonic \ttbar\ decay can be inferred from the momentum imbalance in the event, the allocation of momentum imbalance between the two neutrinos in a dileptonic \ttbar\ decay is unknown a priori.
For this reason, the dileptonic \ttbar\ system is kinematically underconstrained, and mass determination cannot be easily conducted on an event-by-event basis.  Instead, the mass of the parent top quarks in the dileptonic \ttbar system can be extracted from kinematic features over an ensemble of events, with the help of appropriate observables and reconstruction techniques.

The measurement reported in this paper is based on a set of observables that have been proposed specifically for mass reconstruction in underconstrained decay topologies.  These observables include the invariant mass, \mbl, of a \bl\ system, a `stransverse mass' variable, \mttbb, constructed with the $\PQb$ and $\PAQb$ daughters of the \ttbar\ system \cite{mt2_1,mt2_2,minimal_constraints}, and the invariant mass of a \blv\ system, \mblv, where the neutrino momentum is estimated by the \mtt-assisted on-shell (MAOS) reconstruction technique \cite{maos}.
The MAOS reconstruction technique builds on \mtt\ by exploiting the neutrino momenta estimates that are by-products of the \mtt\ algorithm.
The sensitivity of the \mbl, \mttbb, and \mblv\ observables to the value of \mt  is investigated using a Fisher information density method.
Distributions of \mbl\ and \mttbb\ in dileptonic events contain a sharp edge descending to a kinematic endpoint, the location of which is sensitive to the value of \mt.  Recently, masses of the top quark, W boson (\mw), and neutrino (\mnu) were extracted in a simultaneous fit using the endpoints of these distributions in dileptonic \ttbar\ events \cite{endpoints}.
The \mbl, \mttbb, and MAOS \mblv\ observables are described in more detail in Section~\ref{sec:variables}.

One of the dominant sources of systematic uncertainty limiting the precision of this measurement comes from the overall uncertainty in jet energy scale (JES).  To address the JES uncertainty, we introduce a technique that uses the \mbl\ and \mttbb\ observables to determine an overall jet energy scale factor (JSF) simultaneously with the top quark mass, where the JSF is defined as a multiplicative factor scaling the four-vectors of all jets in the event.
Similar techniques have been developed for the all-hadronic and semileptonic \ttbar\ channels, where the jet pair originating from a \PW\ boson decay is used to determine the JSF \cite{cms_comb}.
Because light-quark jets from the \PW\ boson decay are used to calibrate the energy scale of \PQb jets arising from the \cPqt\ and $\overline{\cPqt}$ decays, these methods are sensitive to flavor-dependent uncertainties that emerge from differences in the response of \PQb jets and light-quark jets.
In the method featured here, the JSF is determined in the dileptonic \ttbar\ channel without relying on a \PW\ boson decaying to jets.  Instead, it achieves sensitivity to the JSF through the kinematic differences between \PQb jets, which are subject to JSF scaling, and leptons, which are not.
Because it does not use light quarks from a hadronic W boson decay, this approach is insensitive to flavor-dependent JES uncertainties.

To model the \mbl, \mttbb, and MAOS \mblv\ distribution shapes, we use a Gaussian process (GP) regression technique \cite{rasmussen+williams,bishop}.  This technique is nonparametric, and thus largely model-independent.  It is effective in modeling distribution shapes when no theoretical guidance is available to specify a functional form.  The distribution shapes can conveniently be modeled as functions of multiple variables.  In this analysis, three variables are used: the value of the relevant observable (\mbl, \mttbb, or \mblv), \mt, and the JSF.  The shapes are determined using simulated events generated with seven different values of \mt  ranging from $166.5$ to $178.5\GeV$, and with five values of JSF, ranging from $0.97$ to $1.03$, applied to the jets in each event.  Each shape ultimately models the distributions of the observables together with their evolution in \mt  and in JSF.

\section{The CMS detector}
\label{sec:detector}

The central feature of the CMS apparatus is a superconducting solenoid of 6\unit{m} internal diameter, providing a magnetic field of 3.8\unit{T}. Within the solenoid volume are a silicon pixel and strip tracker, a lead tungstate crystal electromagnetic calorimeter (ECAL), and a brass and scintillator hadron calorimeter (HCAL), each composed of a barrel and two endcap sections.
The tracker has a track-finding efficiency of more than 99\% for muons with transverse momentum $\pt > 1\GeV$ and pseudorapidity $\abs{\eta} < 2.4$.  The ECAL is a fine-grained hermetic calorimeter with quasi-projective geometry, and is distributed in the barrel region of $\abs{\eta}< 1.48$ and in two endcaps that extend up to $\abs{\eta} < 3.0$.  The HCAL barrel and endcaps similarly cover the region $\abs{\eta} < 3.0$.  In addition to the barrel and endcap detectors, CMS has extensive forward calorimetry.  Muons are measured in gas-ionization detectors, which are embedded in the steel flux-return yoke outside of the solenoid.
The silicon tracker and muon systems play a crucial role in the identification of jets originating from the hadronization of \PQb quarks \cite{btagging}.
Events of interest are selected using a two-tiered trigger system~\cite{Khachatryan:2016bia}. The first level, composed of custom hardware processors, uses information from the calorimeters and muon detectors to select events at a rate of around 100\unit{kHz} within a time interval of less than 4\mus. The second level, known as the high-level trigger, consists of a farm of processors running a version of the full event reconstruction software optimized for fast processing, and reduces the event rate to less than 1\unit{kHz} before data storage.
A more detailed description of the CMS detector, together with a definition of the coordinate system used, can be found in Ref.~\cite{Chatrchyan:2008zzk}.

\section{Data sets and event selection}
\label{section:data}

We select dileptonic \ttbar\ events from a data set recorded at $\sqrt{s}=8$\TeV during 2012 corresponding to an integrated luminosity of \luminosity\ \cite{CMS_lumi}.
Events are required to pass one of several triggers that require at least two leptons, $\Pe\Pe$, $\Pe\mu$, or $\mu\mu$, where the leading (higher-\pt) lepton satisfies $\pt > 17\GeV$ and the subleading lepton satisfies $\pt > 8\GeV$.

A particle-flow (PF) algorithm \cite{particle-flow-1,particle-flow-2} is used to reconstruct and identify each individual particle in an event by combining information from various subdetectors of CMS.  Each event is required to have at least one reconstructed collision vertex, with the primary vertex selected as the one containing the largest $\sum\pt^2$ of associated tracks.  Electron candidates are reconstructed by matching a cluster of energy deposits in the ECAL to a reconstructed track \cite{electron}.  They are required to satisfy $\pt > 20\GeV$ and $\abs{\eta} < 2.5$.  Muon candidates are reconstructed in a global fit that combines information from the silicon tracker and muon system \cite{muon}, and must have $\pt > 20\GeV$ and $\abs{\eta} < 2.4$.  A requirement on the relative isolation is imposed inside a cone $\Delta R = \sqrt{\smash[b]{(\Delta\eta)^2 + (\Delta\phi)^2}}$ around each lepton candidate, where $\phi$ is the azimuthal angle in radians.  A parameter $\Irel = \sum{{\pt}_i}/\pt^{\ell}$ is defined, where the sum includes all reconstructed PF candidates inside the cone (excluding the lepton itself), and $\pt^{\ell}$ is the lepton \pt.  Electron (muon) candidates are required to have $\Irel < 0.15$ $(0.2)$ with $\Delta R < 0.3$ $(0.4)$.
Events selected offline are required to contain exactly two such leptons, $\Pe\Pe$, $\Pe\mu$, or $\mu\mu$, with opposite charge.
For events containing an \EE\ or \MM\ pair, contributions from low-mass resonances are suppressed by requiring an invariant mass of the lepton pair $M_{\ell\ell} > 20\GeV$, while contributions from \PZ\ boson decays are suppressed by requiring that $\abs{M_{\PZ} - M_{\ell\ell}} > 15\GeV$, where $M_{\PZ} = 91.2\GeV$ \cite{pdg}.

Hadronic jets are clustered from PF candidates with the infrared and collinear safe anti-\kt algorithm \cite{Cacciari:2008gp}, with a distance parameter $R$ of 0.5, as implemented in the \FASTJET package \cite{fastjet}. The jet momentum is determined as the vectorial sum of all particle momenta in this jet. Corrections to the JES and jet energy resolution (JER) are derived using MC simulation, and are confirmed with measurements of the energy balance in quantum chromodynamics (QCD) dijet, QCD multijet, photon+jet, and Z+jet events~\cite{jes}.  Muons, electrons, and charged hadrons originating from multiple collisions within the same or nearby bunch crossings (pileup), are not included in the jet reconstruction.  Contributions from neutral hadrons originating from pileup are estimated and subtracted from the JES.  Jets originating from the hadronization of \PQb quarks are identified with a combined secondary vertex (CSV) \PQb tagging algorithm \cite{btagging}, combining information from the jet secondary vertex with the impact parameter significances of its constituent tracks.  The algorithm yields a tagging efficiency of approximately $85\%$ and a misidentification rate of $10\%$.  Events are required to contain at least two jets that pass the \PQb tagging algorithm and satisfy $\pt > 30\GeV$ and $\abs{\eta} < 2.5$.  In this analysis, the two jets satisfying these requirements that have the highest CSV discriminator values are referred to as \PQb jets.

The missing transverse momentum vector is defined as $\ptvecmiss = -\sum{\ptvec{}_i}$, where the sum includes all reconstructed PF candidates in an event \cite{met}.  Its magnitude is referred to as \ptmiss.  Corrections to the JES and JER are propagated into \ptmiss, as well as an offset correction that accounts for pileup interactions.  An additional correction mitigates a mild azimuthal dependence, arising from imperfect detector alignment and other effects, which is observed in the reconstructed \ptmiss.  To further suppress contributions from Drell--Yan processes, events containing an \EE\ or \MM\ pair are required to have $\ptmiss > 40\GeV$.

Simulated \ttbar\ signal events are generated with the \MADGRAPH 5.1.5.11 matrix-element generator \cite{madgraph}, combined with M\textsc{ad}S\textsc{pin} to include spin correlations of the top quark decay products \cite{madspin}, \PYTHIA 6.426 with the $\mathrm{Z}2^*$ tune for parton showering \cite{pythia}, and \TAUOLA for the decay of $\tau$ leptons \cite{tauola}.  Parton distribution functions (PDFs) are described by the CTEQ6L1 set \cite{cteq}.  The \ttbar\ signal events are generated with seven different values of \mt  ranging from $166.5$ to $178.5\GeV$.  The contribution from the \PW\ associated single top quark production (t\PW) is simulated with \POWHEG 1.380 \cite{powheg1,powheg2,powheg3,powheg4}, where the value of \mt  is assumed to be $172.5\GeV$.  Background events from \PW+jets and Z+jets production are generated with \MADGRAPH~5.1.3.30, and contributions from \PW\PW, \PW\PZ, $\Z\Z$ processes are simulated with \PYTHIA.  The CMS detector response to the simulated events is modelled with \GEANTfour \cite{geant}.  All background processes are normalized to their predicted cross sections \cite{xsec1,xsec2,xsec3,xsec4,xsec5}.

With the requirements outlined previously, \nevts\ \ttbar\ candidate events are selected in data.  The sample composition is estimated in simulation to be 95\% dileptonic \ttbar, 4\% single top quark, and 1\% other processes including diboson, \PW+jets, and Drell--Yan production, as well as semileptonic and all-hadronic \ttbar.

\section{Observables}
\label{sec:variables}

The observables featured in this study have been developed for physics scenarios where undetected particles, such as neutrinos, carry away a portion of the kinematic information necessary for full event reconstruction.  In the dileptonic \ttbar\ system, distributions in these observables contain endpoints, edges, and peak regions that are sensitive to the top quark mass.
The observables are described in more detail below.

\subsection{The \texorpdfstring{\mbl}{M[b ell]} observable}
\label{sec:mbl}
The \mbl\ observable is defined as
\begin{linenomath*}
  \begin{equation}
    \label{eq:mbl_def}
    \mbl = \sqrt{(p_{\PQb} + p_{\ell})^2},
  \end{equation}
\end{linenomath*}
where $p_{\PQb}$ and $p_{\ell}$ are four-vectors corresponding to a \PQb jet and lepton, respectively.  The \bl\ pairs underlying each value of \mbl\ are chosen out of four possible combinations by an algorithm described below.  The \mbl\ observable contains a kinematic endpoint that occurs when the \PQb jet and lepton are directly back-to-back in the top quark rest frame.  The location of this endpoint, $(\mbl)_{\text{max}}$, is a function of the masses involved in the decay:
\begin{linenomath*}
  \begin{equation}
    \label{eq:mbl_endpnt}
    (\mbl)_{\text{max}} = \frac{\sqrt{(\mt^2-\mw^2)(\mw^2-\mnu^2)}}{\mw}.
  \end{equation}
\end{linenomath*}
With $\mt = 172.5$\GeV, $\mw = 80.4$\GeV \cite{pdg}, and $\mnu = 0$, we have $(\mbl)_{\text{max}} = 152.6$\GeV.  Although this endpoint is a theoretical maximum on the value of \mbl\ at leading order, events are still observed beyond this value due to background contamination, resolution effects, and nonzero particle widths.

\begin{figure}
  \centering
\includegraphics[width=0.49\textwidth]{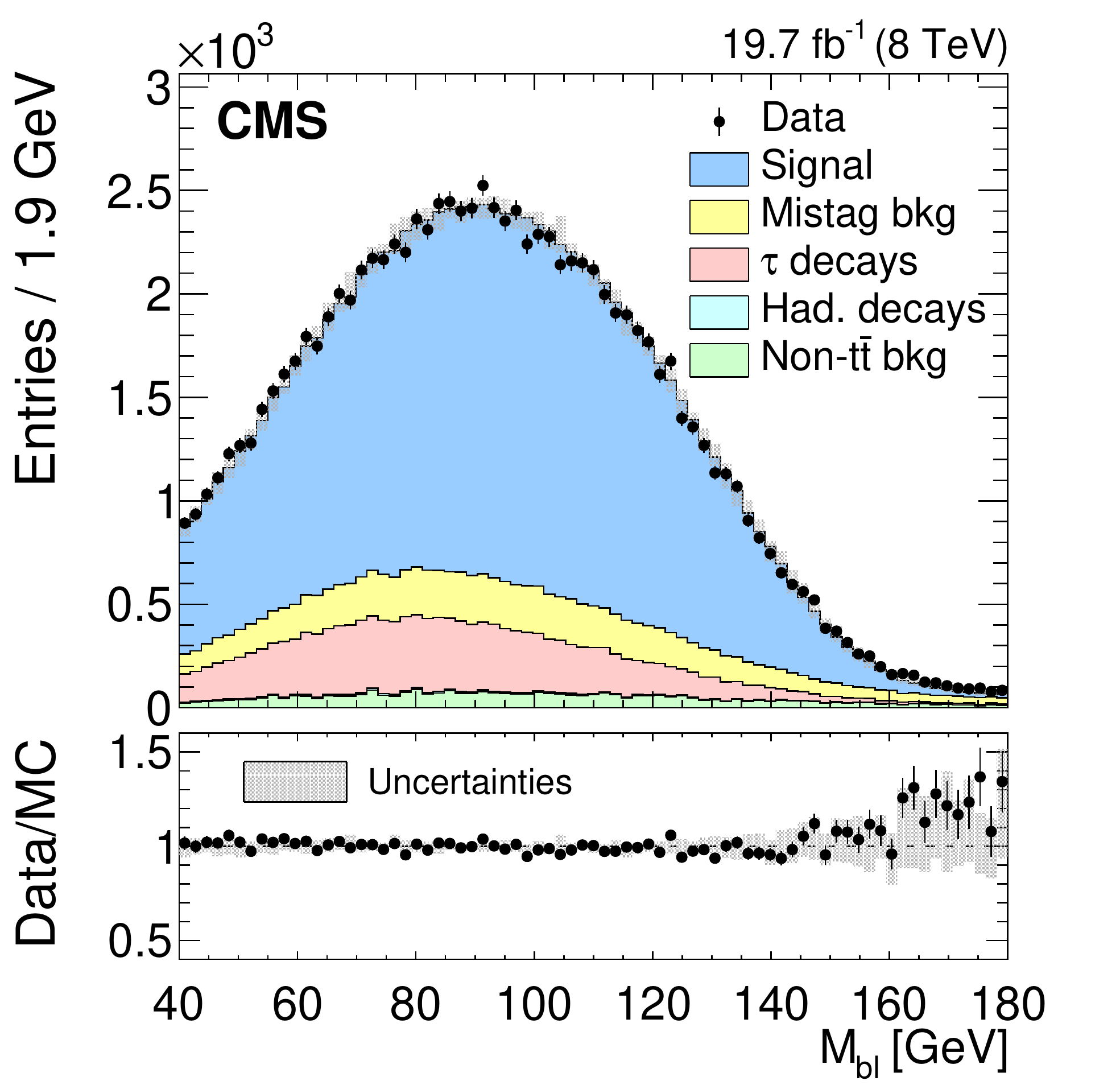}
\includegraphics[width=0.49\textwidth]{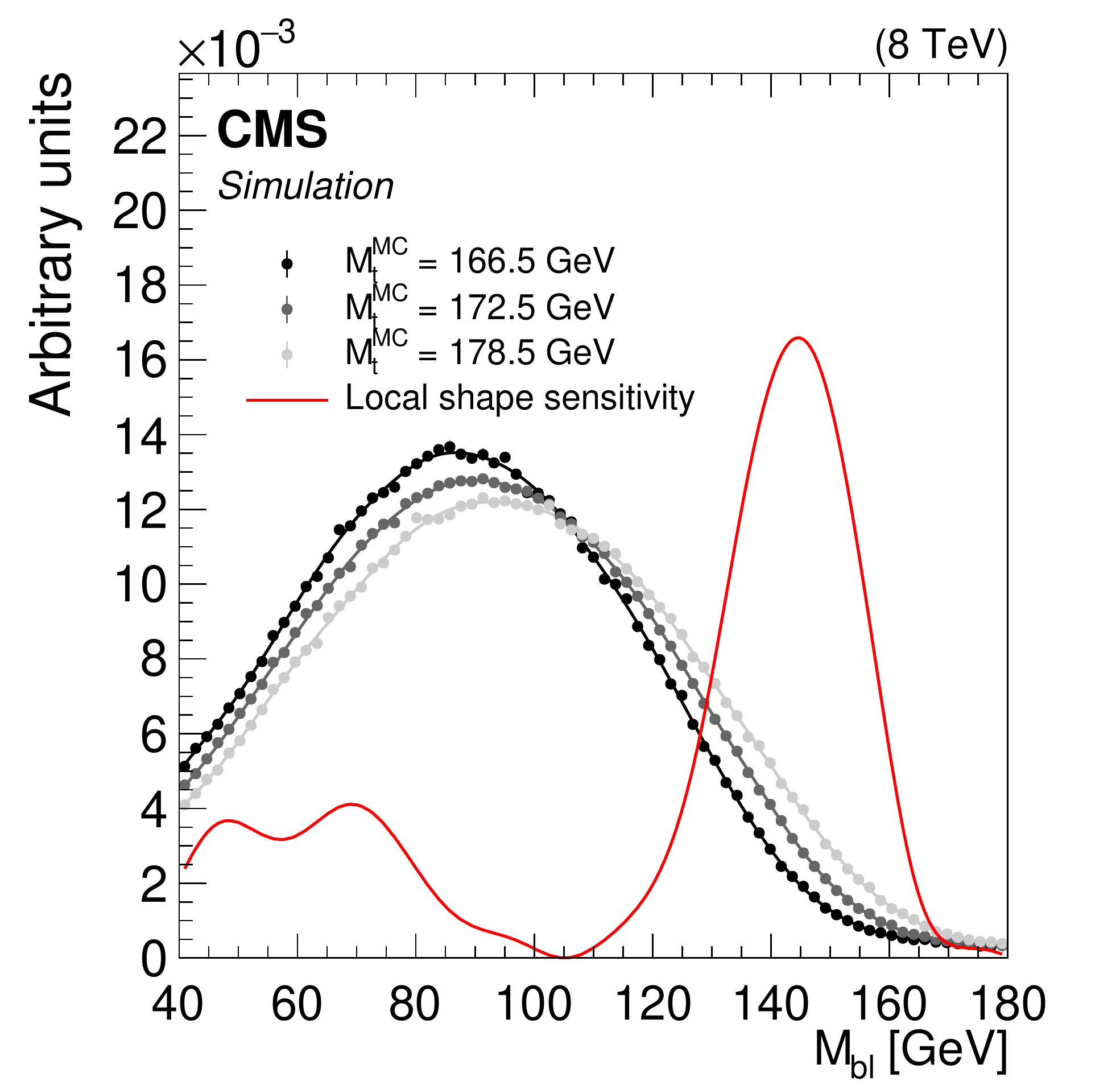}
\caption{(\cmsLLeft) the \mbl\ distribution in data and simulation with $\mtmc = 172.5\GeV$, normalized to the number of events in the 8 TeV data set corresponding to an integrated luminosity of \luminosity.  The lower panel shows the ratio between the data and simulation.  Statistical and systematic uncertainties on the distribution in simulation are represented by the shaded area.  A description of the systematic uncertainties is given in Section~\ref{sec:systematics}.  (\cmsRRight) the \mbl\ distribution shapes in simulation, normalized to unit area, corresponding to three values of \mtmc\ are shown together with the `local shape sensitivity' function, described in Appendix~\ref{sec:sensitivity}.  The \mbl\ distributions include two or three values of \mbl\ for each event.
The distribution shapes are modeled with a GP regression technique, described in Section~\ref{sec:gp}.
}
  \label{fig:dist_mbl}
\end{figure}

The \mbl\ distribution is shown in data and MC simulation in Fig.~\ref{fig:dist_mbl} (\cmsLeft), with a breakdown of signal and background events shown in the simulation.  The `signal' category includes \ttbar\ dilepton decays where both \PQb jets are correctly identified by the \PQb tagging algorithm.  The background categories include: `mistag' dilepton decays where a light quark or gluon jet is incorrectly selected by the \PQb tagging algorithm; `$\tau$ decays' where dilepton events include at least one $\tau$ lepton in the final state subsequently decaying leptonically; and `hadronic decays' that include events where at least one of the top quarks decays hadronically.  The `non-\ttbar\ bkg' category consists of single top quark, diboson, \PW+jets, and Drell--Yan processes.  Events in which a top quark decays through a $\tau$ lepton contain extra neutrinos stemming from the leptonic $\tau$ decay.  Although the extra neutrinos cause a small distortion to the kinematic distributions, these events still contribute to the sensitivity of the measurement.

The sensitivity of the \mbl\ observable to the value of \mt  is demonstrated in Fig.~\ref{fig:dist_mbl} (\cmsRight), where \mbl\ shapes corresponding to three values of the top quark mass in MC simulation (\mtmc) are shown.
The variation between these shapes reveals regions of the \mbl\ distribution that are sensitive to the value of \mt, such as the edges to the left and right of the \mbl\ peak, and regions that are not sensitive, such as the stationary point where the three shapes intersect.
To provide a quantitative description of these effects, we introduce a `local shape sensitivity' function, also known as the Fisher information density, shown in Figs.~\ref{fig:dist_mbl}, \ref{fig:dist_mt2}, and \ref{fig:dist_maos}.  This function conveys the sensitivity of an observable at a specific point on its shape.
For the \mbl\ observable, the local shape sensitivity function peaks near the kinematic endpoint ($\mbl \sim 150\GeV$), and has a zero value at the stationary point ($\mbl \sim 105\GeV$).
The integral of this function over its range is proportional to $1/\sigma_{\mt}^2$, where $\sigma_{\mt}$ is the statistical uncertainty on a measurement of \mt.
A full description of the local shape sensitivity function is given in Appendix~\ref{sec:sensitivity}.

\subsubsection*{\texorpdfstring{\PQb}{b} jet and lepton combinatorics}
\label{sec:mbl_comb}
The two \PQb jets and two leptons stemming from each \ttbar\ decay give rise to a two-fold matching ambiguity, with two correct and two incorrect \bl\ pairings possible in each event.  Pairings in which the \PQb jet and lepton emerge from different top quarks do not necessarily obey the upper bound described in Eq.~\eqref{eq:mbl_endpnt}, and thus do not have a clean kinematic endpoint in \mbl.  Although a priori it is experimentally difficult to distinguish between correct and incorrect pairings, one possible approach is to select the smallest two \mbl\ values in each event.  This way, the kinematic endpoint of the distribution is preserved -- even if the smallest two \mbl\ values do not correspond to the correct pairings, they are guaranteed to fall below the correct pairings, which do respect the endpoint.  In this analysis, we employ a slightly more sophisticated matching technique, introduced in Ref.~\cite{endpoints}, where either two or three \bl\ pairs are selected in each event.

By selecting either two or three \bl\ pairs in each event, the technique employed in this analysis has the benefit of increased statistical power, while preserving the kinematic endpoint of \mbl.
Although they are not necessarily the correct pairs, the corresponding \mbl\ values are guaranteed by construction to be less than or equal to those of the correct pairs.  The matching technique is based on the following prescription:
\begin{enumerate}
  \item match each \PQb jet with the lepton that produces the lower \mbl\ value;
  \item match each lepton with the \PQb jet that produces the lower \mbl\ value.
\end{enumerate}
This recipe produces either two or three values of \mbl.  In the latter case, two different leptons may be successfully paired with the same \PQb jet, and vice versa.  Such a configuration highlights the difference between this recipe and the simpler approach of choosing the smallest two values of \mbl, which do not necessarily incorporate both \PQb jets and both leptons in the event.  For example, this could occur if both \PQb jets are matched to a single lepton.  In these cases, the next largest \mbl\ value is also needed to ensure both \PQb jets and both leptons from the event are used.

\subsection{The \texorpdfstring{\mtt}{M[T2]} observable}
\label{subsec:mt2}
The \mtt\ `stransverse mass' observable \cite{mt2_1,mt2_2} is based on the transverse mass, \mtr.  The transverse mass of the W boson in a $\PW\to \ell\nu$ decay is given by
\begin{linenomath*}
  \begin{equation}
    \label{eq:transverse_mass}
    \mtr = \sqrt{ m_{\ell}^2 + m_{\nu}^2 + 2(E_{\mathrm{T}\ell}E_{\mathrm{T}\nu} - \vec{p}_{\mathrm{T}\ell}\cdot\vec{p}_{\mathrm{T}\nu}) }\, ,
  \end{equation}
\end{linenomath*}
where $E_{\mathrm{T}x}^2 = m_x^2 + \ptt$ for $x\in\{\ell,\nu\}$, $m_x$ is the particle mass, and $\vec{p}_{\mathrm{T}x}$ is the particle momentum projected onto the plane perpendicular to the beams.  This quantity exhibits a kinematic endpoint at the parent mass, \mw, which occurs in configurations when both the lepton and neutrino momenta lie entirely in the transverse plane (up to a common longitudinal boost).

The dileptonic \ttbar\ system has two layers of decays, with $\cPqt\to \PW\PQb$ in the first step followed by $\PW\to\ell\nu$ in the second.  The result is an event topology with two identical branches, $\cPqt\to \PQb\ell^+\nu$ and $\cPaqt \to \PAQb\ell^-\PAGn$, each with a visible (\bl) and invisible ($\nu$) component.  In this case, one value of \mtr\ can be computed for each branch.  The invisible particle momentum associated with each branch, however, is not known.  While for a semileptonic \ttbar\ decay, with only one $\PW\to \ell\nu$ decay, the neutrino \ptvec\ is estimated from the \ptvecmiss\ in the event, a dileptonic \ttbar\ decay includes two neutrinos, for which the allocation of \ptvecmiss\ between them is unknown.

The \mtt\ observable is an extension of \mtr\ for a system with two identical decay branches, `a' and `b', such as those in the dileptonic \ttbar\ system.  Here, the invisible particle momenta, \pta\ and \ptb, must add up to the total \ptvecmiss.  The strategy of \mtt\ is to impose this constraint on the invisible particle momenta, while also performing a minimization in order to preserve the kinematic endpoint of \mtr.  For a general event with a symmetric decay topology, \mtt\ is defined as
\begin{linenomath*}
  \begin{equation}
    \label{eq:mt2}
    \mtt = \min_{\pta+\ptb=\ptvecmiss}{[\max\{\mtr^{\mathrm{a}},\mtr^{\mathrm{b}}\}]},
  \end{equation}
\end{linenomath*}
where $\mtr^{\mathrm{a}}$ and $\mtr^{\mathrm{b}}$ correspond to the two decay branches.  If the invisible particle mass is known, it can be incorporated into the \mtt\ calculation as well, yielding an endpoint at the parent particle mass.  Although the final values of \pta\ and \ptb\ are typically treated as intermediate quantities in the \mtt\ algorithm, they are employed as neutrino \ptvec\ estimates in the MAOS reconstruction technique described in Section~\ref{subsec:maos}.

\subsubsection*{The \texorpdfstring{\mtt}{M[T2]} subsystems}
\label{subsubsec:mt2sub}

In the \ttbar\ system, there are several ways in which \mtt\ can be computed, depending on how the decay products are grouped together.  The \mtt\ algorithm classifies them into three categories: upstream, visible, and child particles \cite{subsystem_mt2}. The child particles are those at the end of the decay chain that are unobservable or simply treated as unobservable.  In the latter case, the child particle momenta are added to the \ptvecmiss\ vector.  The visible particles are those whose \ptvec\ values are measured and used in the calculations; and the upstream particles are those from further up in the decay chain, including any initial-state radiation (ISR) accompanying the hard collision.

\begin{figure}
  \centering
  \includegraphics[width=0.5\textwidth]{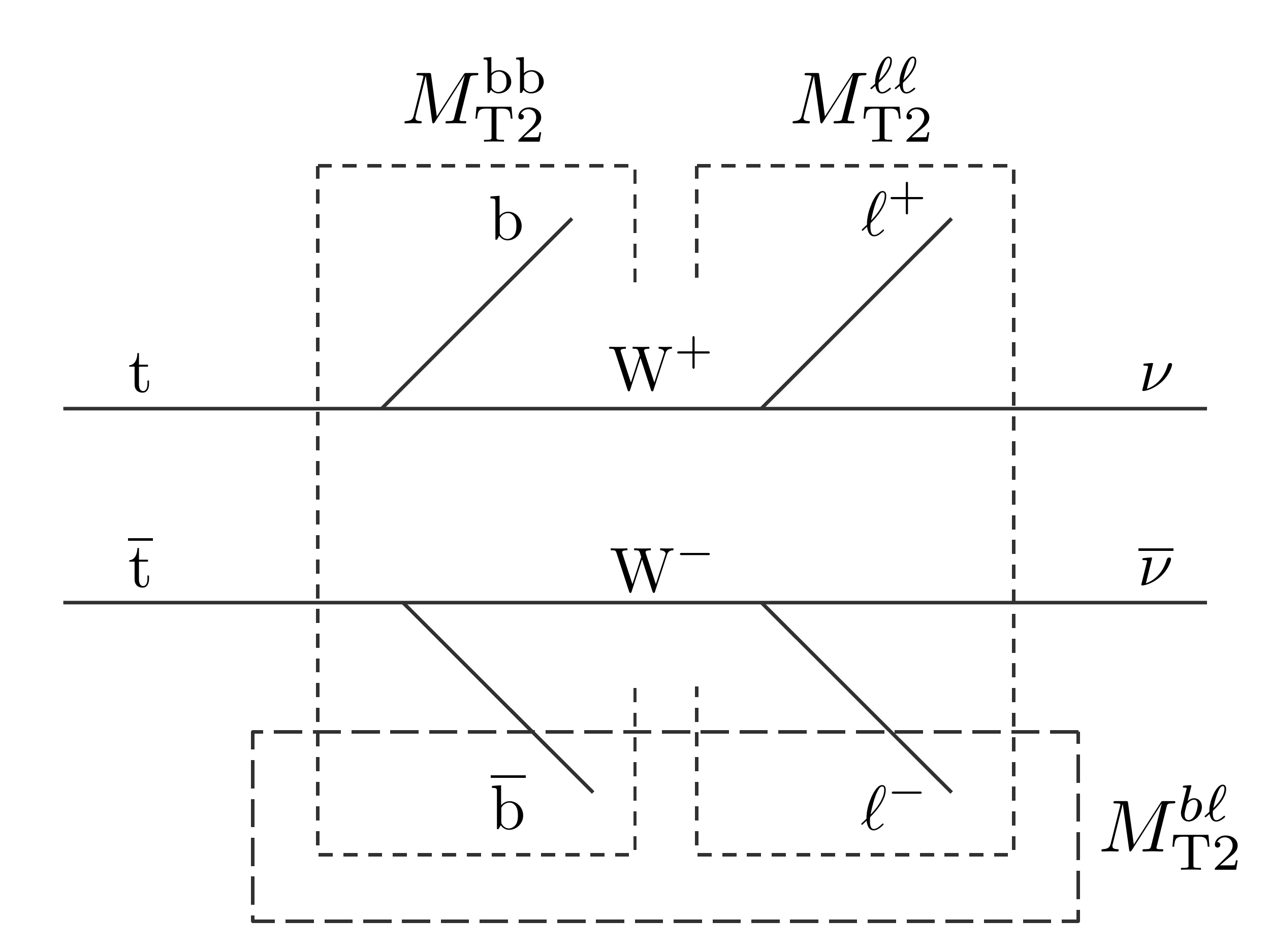}
  \caption{The \mtt\ subsystems in the dileptonic \ttbar\ event topology.}
  \label{fig:mt2_subsystems}
\end{figure}

In general, the child, visible, and upstream particles may actually be collections of objects, creating three possible subsystems in the dileptonic \ttbar\ event topology.  These subsystems are illustrated in Fig.~\ref{fig:mt2_subsystems}.  For simplicity, we refer to the corresponding \mtt\ observables as \mttbb, \mttll, and \mttbl, where:
\begin{itemize}
  \item The \mttllb\ \textbf{observable} uses the two leptons as visible particles, treating the neutrinos as invisible child particles, and combining the \PQb jets with all other upstream particles in the event.
  \item The \mttbbb\ \textbf{observable} uses the \PQb jets as visible particles, and treats the W bosons as child particles, ignoring the fact that their charged daughter leptons are indeed observable.  It considers only ISR jets as generators of upstream momentum.
  \item The \mttblb\ \textbf{observable} combines the \PQb jet and the lepton to form a single visible system, and takes the neutrinos as the invisible particles.  A two-fold matching ambiguity results from the matching of \PQb jets to leptons in each event.  In order to preserve the kinematic endpoint of the \mttbl\ distribution, the $\PQb\ell$ pair with the smallest value of \mttbl\ is used in each event.
\end{itemize}
These observables are identical, respectively, to $\mtt^{(2,2,1)}$, $\mtt^{(2,1,0)}$, $\mtt^{(2,2,0)}$ of Ref.~\cite{subsystem_mt2}, and $\mu_{\mathrm{b}\mathrm{b}}$, $\mu_{\ell\ell}$, $\mu_{\mathrm{b}\ell}$ of Ref.~\cite{endpoints}.

\begin{figure}[htbp]
  \centering
  \includegraphics[width=0.49\textwidth]{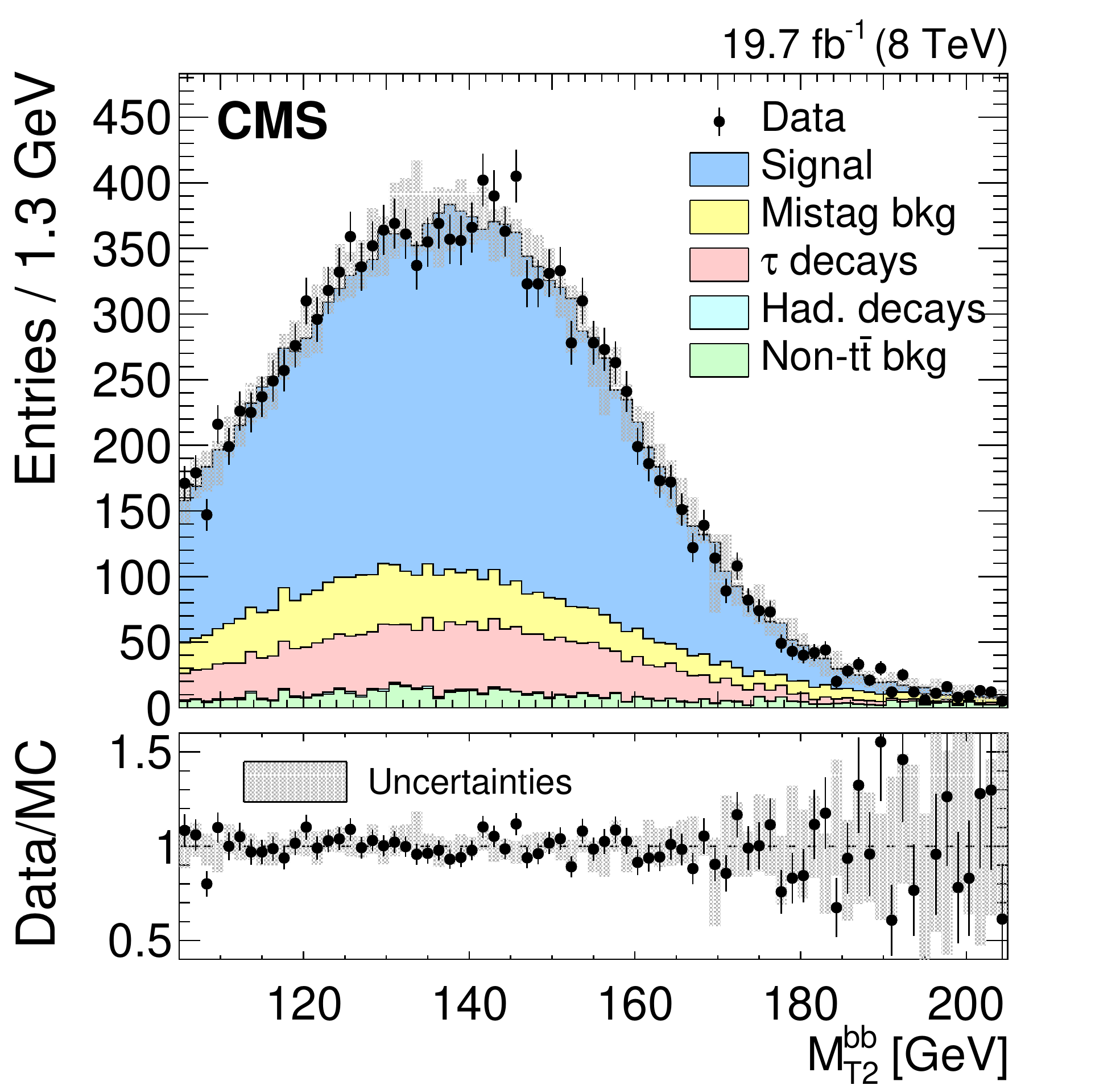}
  \includegraphics[width=0.49\textwidth]{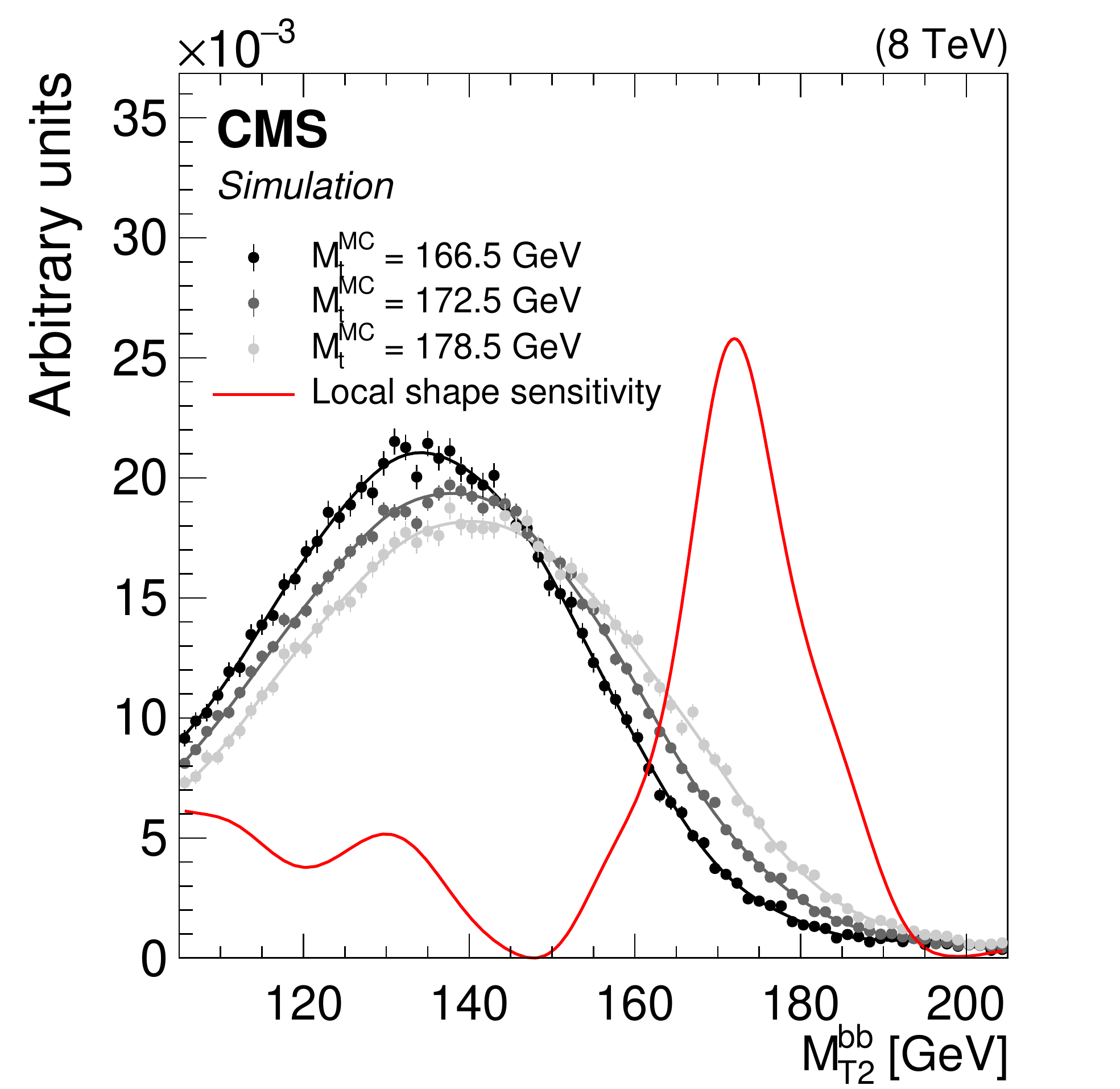}
  \caption{ Following the conventions of Fig.~\ref{fig:dist_mbl}, shown are the (\cmsLeft) \mttbb\ distribution in data and simulation with $\mtmc = 172.5\GeV$, and (\cmsRight) \mttbb\ distribution shapes in simulation corresponding to three values of \mtmc, along with the `local shape sensitivity' function.  The \mttbb\ distributions include one value of \mttbb\ for each event if it satisfies the kinematic requirement outlined in Section~\ref{subsubsec:mt2sub}.}
  \label{fig:dist_mt2}
\end{figure}

The subsystem observable \mttbb\ is employed in this study to complement the observable \mbl.  The \mttbb\ observable contains an endpoint at the value of \mt, and can be combined with \mbl\ to mitigate uncertainties due to the JES.  This feature is discussed further in Section~\ref{sec:jes}.  The distribution of \mttbb\ and its sensitivity to the value of \mt  are shown in Fig.~\ref{fig:dist_mt2}.  Although \mttll\ is not directly sensitive to \mt, the neutrino \ptvec\ estimates that are a by-product of its computation are used as an input into the MAOS \mblv\ reconstruction technique described in Section~\ref{subsec:maos}.

The \mttbb\ distribution employed in this analysis includes a kinematic requirement on the upstream momentum, defined as $\ptupst = \sum_{\rm reco}{\ptvec{}_i} - \sum_{\text{b jets}}{\ptvec{}_i} - \sum_{\text{leptons}}{\ptvec{}_i}$, where the sums are conducted over all reconstructed PF candidates, \PQb jets, and leptons in each event, respectively.
The direction of \ptupst\ is required to lie outside the opening angle between the two \PQb jet \ptvec\ vectors in the event.  This requirement primarily impacts events at low values of \mttbb, and its effect on the statistical sensitivity of the observable is small.

\subsection{The MAOS \texorpdfstring{\mblv}{M[b ell nu]} observable}
\label{subsec:maos}

The MAOS reconstruction technique employed in this analysis is based on the subsystem observable \mttll.  In the \mttll\ algorithm, an \mtr\ variable, defined in Eq.~\eqref{eq:transverse_mass}, is constructed from the $\ell^+\nu$ and $\ell^-\overline{\nu}$ pairs corresponding to each of the \ttbar\ decay branches.  Because the values of neutrino \ptvec\ are unknown, a minimization is conducted in Eq.~\eqref{eq:mt2} over possible values consistent with the measured \ptvecmiss\ in each event.

The MAOS technique employs the neutrino \ptvec values that are determined by the \mttll\ minimization to construct full \blv\ invariant mass estimates corresponding to each of the \ttbar decay branches.
Given the neutrino \ptvec\ values, the remaining $z$-components of their momenta are obtained by enforcing the W mass on-shell requirement \cite{pdg}
\begin{linenomath*}
  \begin{equation}
    \label{eq:maos_wmass}
    M(\ell^+\nu) = M(\ell^-\overline{\nu})= \mw = 80.4\GeV.
  \end{equation}
\end{linenomath*}
This yields a longitudinal momentum for each neutrino given by
\begin{linenomath*}
  \begin{equation}
    \label{eq:maos_pz}
    p_{z\nu} = \frac{1}{E_{\mathrm{T}\ell}^2} \left[ p_{z\ell} A \pm \sqrt{ p_{z\ell}^2 + E_{\mathrm{T}\ell}^2} \sqrt{A^2 - (E_{\mathrm{T}\ell} E_{\mathrm{T}\nu})^2} \right],
  \end{equation}
\end{linenomath*}
where $A = \frac{1}{2}(\mw^2+\mnu^2+M_{\ell}^2) + \ptvec{}_{\ell} \cdot \ptvec{}_{\nu}$ \cite{maos}.
Given these estimates for the neutrino three-momenta together with $\mnu=0$, we have the required four vectors to construct an \mblv\ invariant mass corresponding to the decay products of each top quark.

\begin{figure}
  \centering
  \includegraphics[width=0.49\textwidth]{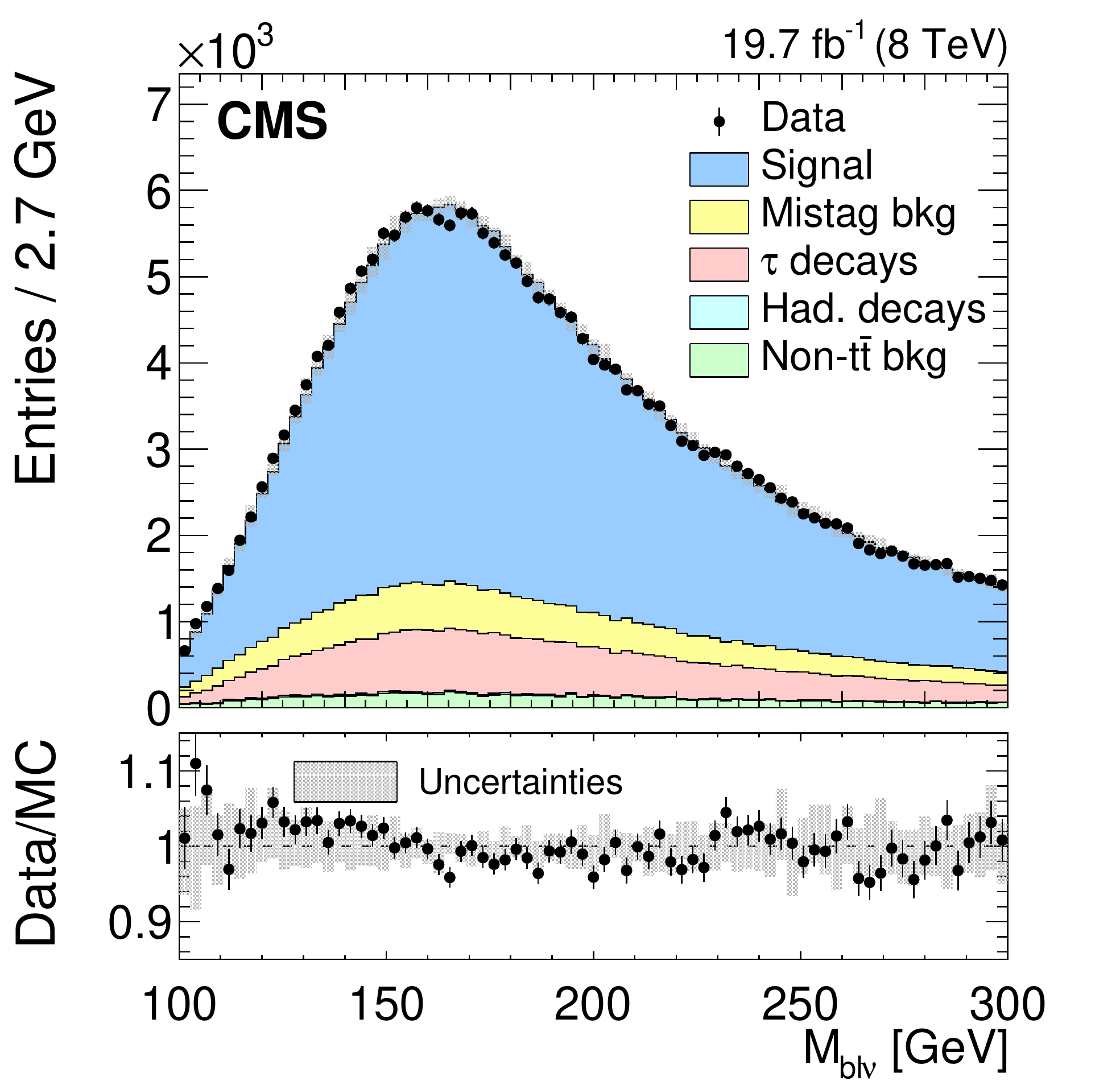}
  \includegraphics[width=0.49\textwidth]{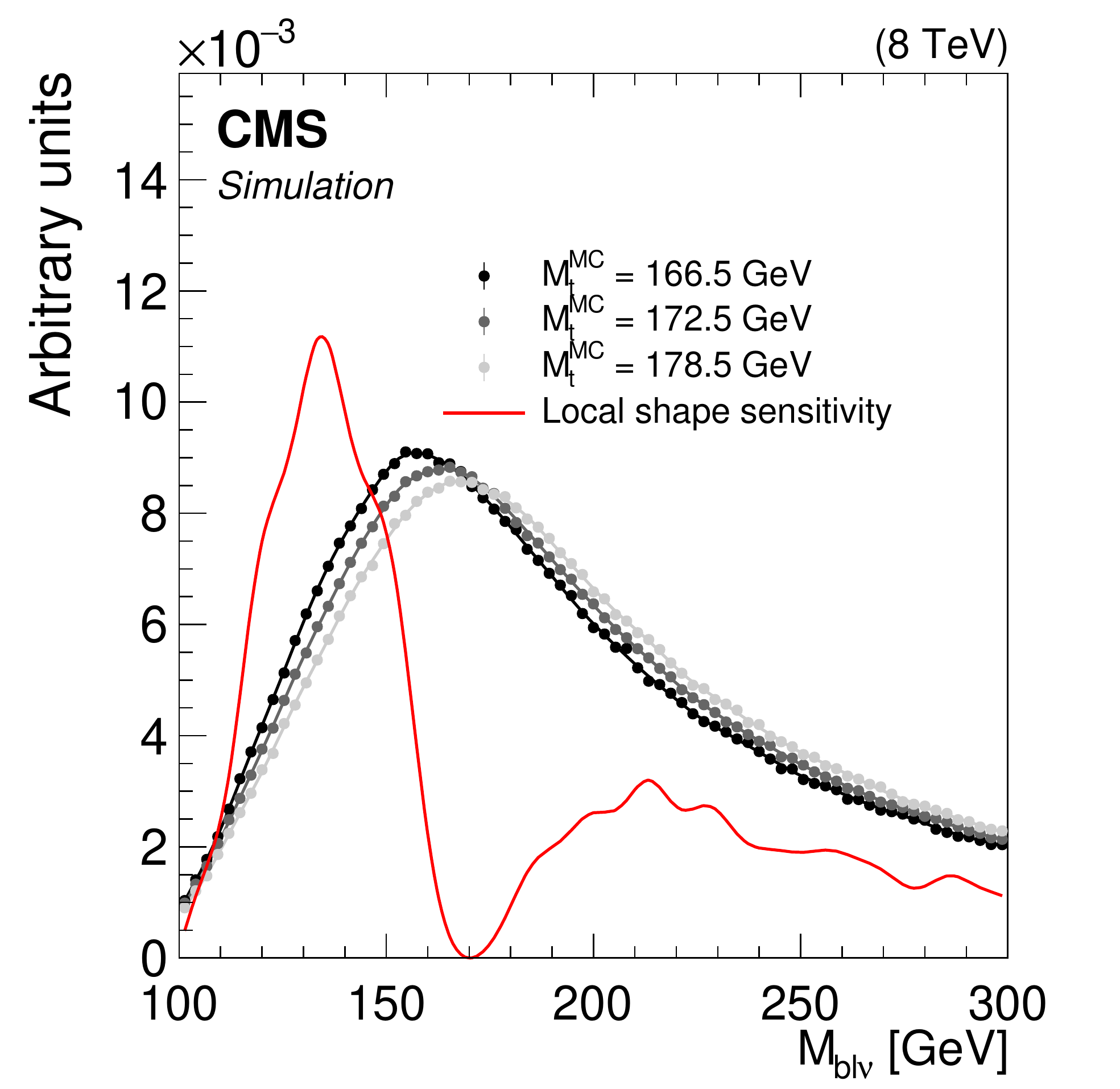}
\caption{
   Following the conventions of Fig.~\ref{fig:dist_mbl}, shown are the (\cmsLeft) MAOS \mblv\ distribution in data and simulation with $\mtmc = 172.5\GeV$, and (\cmsRight) the MAOS \mblv\ distribution shapes in simulation corresponding to three values of \mtmc, along with the `local shape sensitivity' function.  The MAOS \mblv\ distributions include up to eight values of \mblv\ for each event.}
  \label{fig:dist_maos}
\end{figure}

The quadratic equations in Eq.~\eqref{eq:maos_pz} underlying the W mass on-shell requirement provide up to two solutions for each value of $p_{z\nu}$, yielding a two-fold ambiguity for each neutrino momentum.
In addition, there is a two-fold ambiguity resulting from the matching of \PQb jets to $\ell\nu$ pairs in the construction of \blv\ invariant masses.
No matching ambiguity exists between leptons and neutrinos, since the $\ell^+\nu$ and $\ell^-\overline{\nu}$ pairs have been fixed by the \mttll\ algorithm.
The combined four-fold ambiguity, along with the two top quark decays in each event, gives up to eight possible values of \mblv.  In the measurement, all of the available values are used: for each $\ell\nu$ pair, this includes up to two neutrino $p_{z\nu}$ solutions, and two $\PQb$-$\ell\nu$ matches.
The distribution of MAOS \mblv\ and its sensitivity to the value of \mt  are shown in Fig.~\ref{fig:dist_maos}.

\section{Simultaneous determination of \mt\ and JSF}
\label{sec:jes}

To mitigate the impact of JES uncertainties on the precision of this measurement, we introduce a technique that allows a JSF parameter to be fit simultaneously with \mt.  The JSF is a constant multiplicative factor that calibrates the overall energy scale of reconstructed jets.  It is applied in addition to the standard JES calibration, which corrects the jet response as a function of \pt and $\eta$.  The dominant component of uncertainty in the JES calibration can be attributed to a global factor in jet response, which is captured in the JSF.

The challenge in determining the JSF simultaneously with \mt  stems from the large degree of correlation between these parameters.  In the top quark decay, $\PQt\to\blv$, the JSF directly affects the momentum of the \PQb jet, and indirectly, the inferred momentum of the neutrino, by scaling all jets entering the \ptmiss\ sum.  The \mt  parameter affects the momenta of these two particles in addition to the lepton produced in the top quark decay.  In the context of observables and distribution shapes, variations in the \mt  and JSF parameters cause shape changes that are difficult to distinguish.  For this reason, a shape-based analysis using a single observable can be implemented to determine either \mt  or JSF, but not both simultaneously.

\begin{figure}
  \centering
  \includegraphics[width=0.49\textwidth]{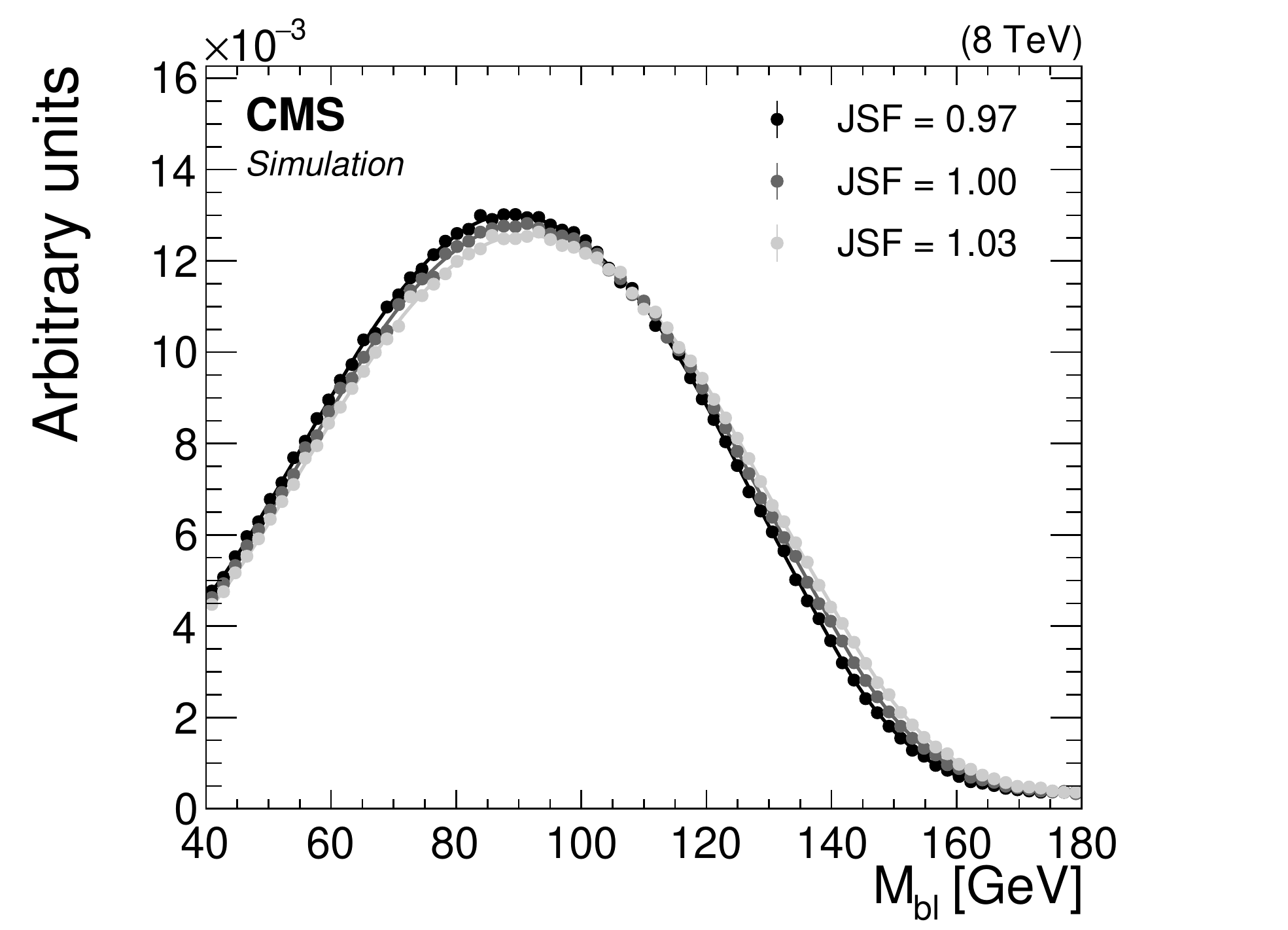}
  \includegraphics[width=0.49\textwidth]{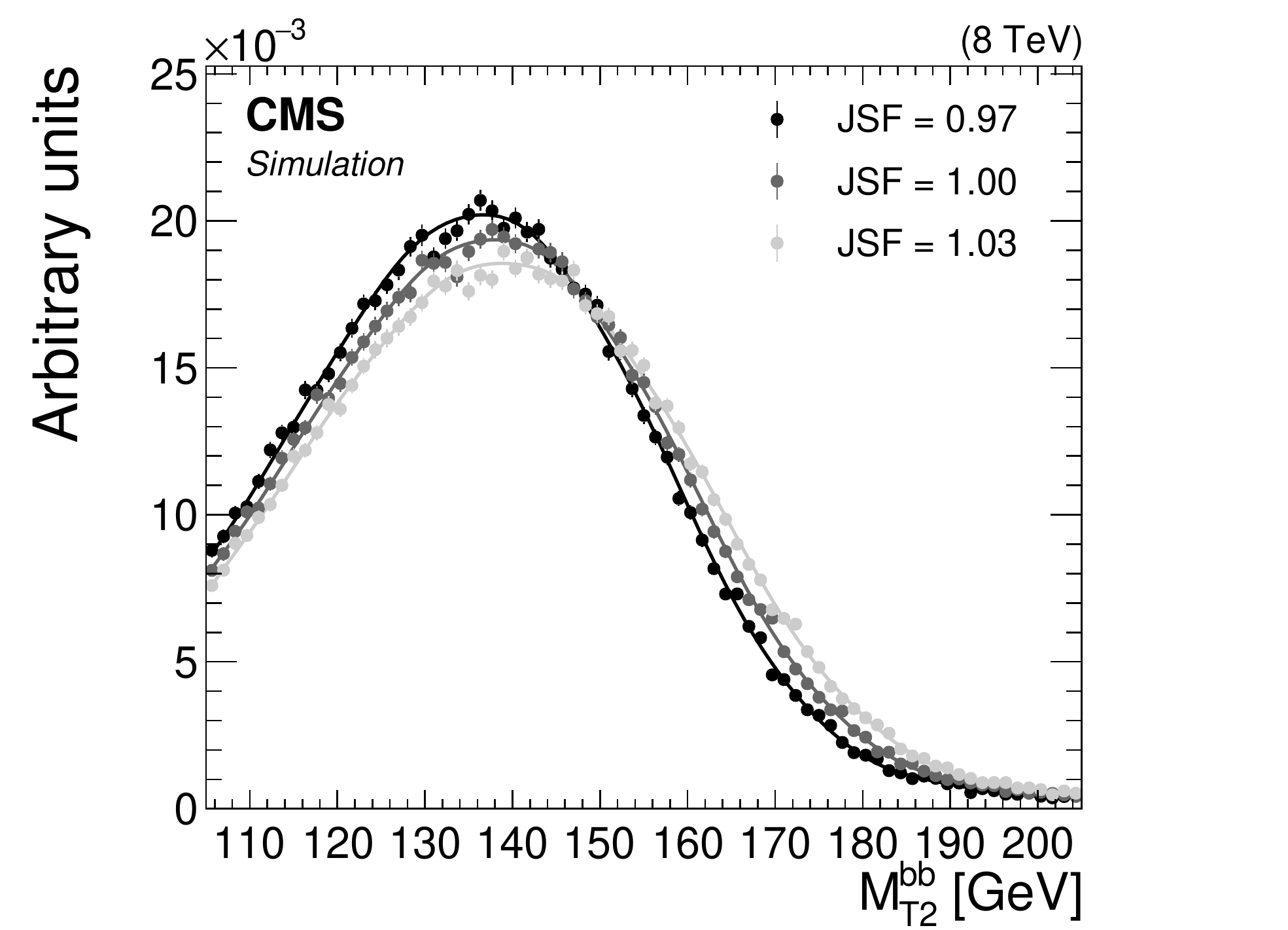}
  \caption{The (\cmsLeft) \mbl\ and (\cmsRight) \mttbb\ distributions in simulation with $\mt = 172.5$\GeV for several values of JSF.  Two or three values are included in the \mbl\ distribution for each event, and one value is included in the \mttbb\ distribution if it satisfies the kinematic requirement outlined in Section~\ref{subsubsec:mt2sub}.  The distributions are normalized to unit area.  The three curves corresponding to each of the \mbl\ and \mttbb\ distributions are obtained using a GP regression technique described in Section~\ref{sec:gp}.}
  \label{fig:mbl_mt2_jsf}
\end{figure}

To determine the \mt  and JSF parameters simultaneously, we construct a likelihood function that contains two distributions corresponding to the \mbl\ and \mttbb\ observables.  In this configuration, variations in the parameters produce shifts in each individual distribution.  They also create a relative shift between the distributions that provides the additional constraint needed for a simultaneous fit of \mt  and JSF.  The dependence of the \mbl\ and \mttbb\ distribution shapes on \mt  is shown in Figs.~\ref{fig:dist_mbl} and \ref{fig:dist_mt2}, and their dependence on the JSF is shown in Fig.~\ref{fig:mbl_mt2_jsf}.  The difference in response between the \mbl\ and \mttbb\ shapes to the JSF parameter is rooted in the reconstructed objects underlying the \mbl\ and \mttbb\ observables -- while each value of \mbl\ uses one \PQb jet and one lepton, each value of \mttbb\ uses two \PQb jets and no leptons for the visible system.  Thus, \mttbb\ exhibits a stronger dependence on the JSF.  The likelihood fit used in this measurement is described in more detail in Section~\ref{sec:fit}.

\section{Gaussian processes for shape estimation}
\label{sec:gp}

In this analysis, the \mbl, \mttbb, and \mblv\ distribution shapes are modeled with a GP regression technique that has two main advantages over other commonly-used shape estimate methods.  First, the GP shape is nonparametric, determined only by a set of training points and hyperparameters that regulate smoothing; and second, it can be easily trained as a function of several variables simultaneously.  The latter feature allows one to capture the smooth evolution of the distribution shapes as the \mt  and JSF parameters are varied.
A detailed introduction to GPs can be found in Refs.~\cite{rasmussen+williams,bishop}.
Here, we give a brief overview of the GP regression technique, with further discussion provided in Appendix~\ref{sec:gp_appendix}.

The likelihood fit described in Section~\ref{sec:fit} uses distribution shapes of the form $\fa(x|\mt,\jsf)$, where $x$ is the value of an observable (\mbl, \mttbb, or \mblv), and \mt  and \jsf\ are free parameters in the fit.  The shapes \fa are shown in Figs.~\ref{fig:dist_mbl}, \ref{fig:dist_mt2}, and \ref{fig:dist_maos} for each observable, where the free parameters are set to $\mt=166.5,172.5,$ or $178.5\GeV$ and $\jsf=1$.  In Fig.~\ref{fig:mbl_mt2_jsf}, shapes corresponding to the \mbl\ and \mttbb\ observables are shown with the free parameters set to $\mt=172.5\GeV$ and $\jsf = 0.97, 1.00,$ or $1.03$.  In the figures, these shapes are represented as functions of a single variable (the observable $x$) with \mt  and \jsf\ fixed.  In GP regression, however, each shape is treated as a function of all three quantities ($x$, \mt, and \jsf), and can be described as a probability density in three dimensions.

Each GP shape is trained using binned distributions of the observable $x$ in MC simulation.  For each observable, $35$ binned distributions are used, corresponding to seven values of \mtmc\ ranging from $166.5$ to $178.5$\GeV and five values of \jsf\ ranging from $0.97$ to $1.03$.  Each distribution has 75 bins in $x$, yielding a total of $2625$ training points at which the value of \fa is known and used as an input into the GP regression process.
Each training point is specified by its values of $x$, \mt, and \jsf.
The GP regression technique interpolates between the discrete values of $x$, \mt, and \jsf\ covered by these training points to provide a shape that is smooth over its range.  The smoothness properties of each shape are determined by a kernel function that is set by the analyzer.  The GP shapes in this analysis correspond to the kernel function given in Eq.~\ref{eq:gp_kernel} of Appendix~\ref{sec:gp_appendix}.

The binned distributions used to construct each GP shape are normalized to unity.  However, the normalization of the GP shape itself may deviate slightly from unity due to minor imperfections in shape modeling.  To mitigate this effect, the GP shape normalization is recomputed for each value of \mt  and JSF at which the shape is evaluated.  In a likelihood fit, the normalization is recomputed for every variation of the fit parameters.

\section{Fit strategy}
\label{sec:fit}

This measurement employs an unbinned maximum-likelihood fit using the \mbl, \mttbb, and MAOS \mblv\ observables described in Section~\ref{sec:variables}, along with the GP shape estimate technique described in Section~\ref{sec:gp}.  The MC samples used to train the GP shapes include the \ttbar\ signal and background processes described in Section~\ref{section:data}.

The likelihood constructed from a single observable, $x$, is given by:
\begin{linenomath*}
  \begin{equation}
    \label{eq:L}
    \mathcal{L}^x(\mt, \jsf) = \prod_i{\fa(x_i | \mt, \jsf)}.
  \end{equation}
\end{linenomath*}
Here, the distribution shape \fa depends on the value of the free parameters \mt  and \jsf, and expresses the likelihood of drawing some event $i$ where the value of the observable is $x_i$.  It is normalized to unity over its range for all values of \mt  and \jsf.  The parameters \mt  and \jsf\ are varied in the fit to maximize the value of the likelihood.

A likelihood containing two observables, $x_1$ and $x_2$, is constructed as a product of individual likelihoods:
\begin{linenomath*}
  \begin{equation}
    \label{eq:Lprod}
    \begin{aligned}
      \mathcal{L}(\mt,\jsf) &= \mathcal{L}^{x_1}(\mt, \jsf)\,\mathcal{L}^{x_2}(\mt, \jsf) \\
                            &= \prod_i{\fa(x_{1i} | \mt, \jsf) \, \fa(x_{2i} | \mt, \jsf)}.
    \end{aligned}
  \end{equation}
\end{linenomath*}
This analysis employs three different versions of the likelihood fit:
\begin{enumerate}
  \item the \textbf{1D fit} uses the \mbl\ and \mttbb\ observables to determine \mt, and JSF is constrained to be unity;
  \item the \textbf{2D fit} also uses \mbl\ and \mttbb\, but imposes no constraint on the JSF and determines \mt  and JSF simultaneously;
  \item the \textbf{MAOS fit} uses the \mttbb\ and \mblv\ observables to determine \mt, and JSF is constrained to be unity.
\end{enumerate}
Among these versions, the 1D fit provides the best precision on the value of \mt.  The 2D fit mitigates the JES uncertainties, which are the largest source of systematic error in the 1D approach.  The MAOS fit is expected to yield results similar to the 1D fit, and is presented as a viable alternative that substitutes the \mbl\ observable for MAOS \mblv.  The best overall precision on \mt  is given by a combination of the 1D and 2D fits, which is discussed below.  The fit results are discussed in Section~\ref{sec:results}.

The central value and statistical uncertainty on \mt  and JSF are determined using the bootstrapping technique \cite{bootstrap}.
This method is based on \PEs rather than the shape of the total likelihood defined in Eq.~\eqref{eq:Lprod} near its maximum, and thus mitigates the effects of correlation between the two observables, $x_1$ and $x_2$, in the likelihood.  The technique also mitigates possible correlations within the \mbl\ and \mblv\ observables when multiple values of the observable occur in a single event.
The bootstrapping technique is primarily relevant for statistical uncertainty determination, which may otherwise be affected by correlations in the likelihood.  The technique has a negligible impact on the central values of \mt  and JSF.
The bootstrap \PEs are constructed by resampling the full data set with replacement, where the size of each \PE is fixed to have the number of events in data (\nevts events).  Events are selected at random from the full data set, so that a particular event has the same probability of being chosen at any stage during the sampling process.  In this procedure, a single event may be selected more than once for any given \PE.  In data, all events have an equal probability to be selected.
In simulation, the probability of selecting a particular event is proportional to its weight, containing the relevant cross sections, as well as corrections for MC modeling and object reconstruction efficiencies.

\begin{figure*}
  \centering
  \includegraphics[width=\cmsFigMultiWidth]{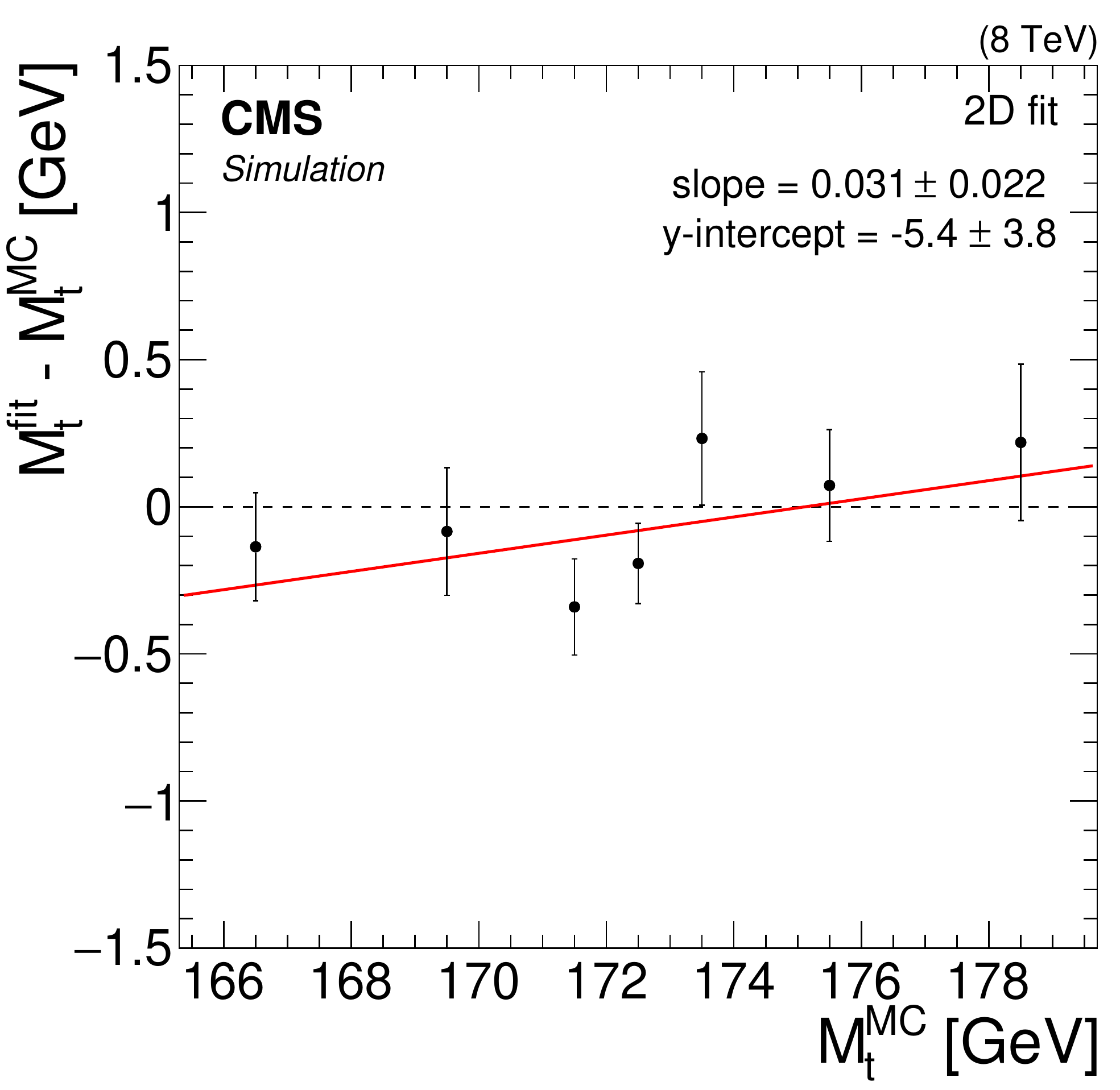}
  \includegraphics[width=\cmsFigMultiWidth]{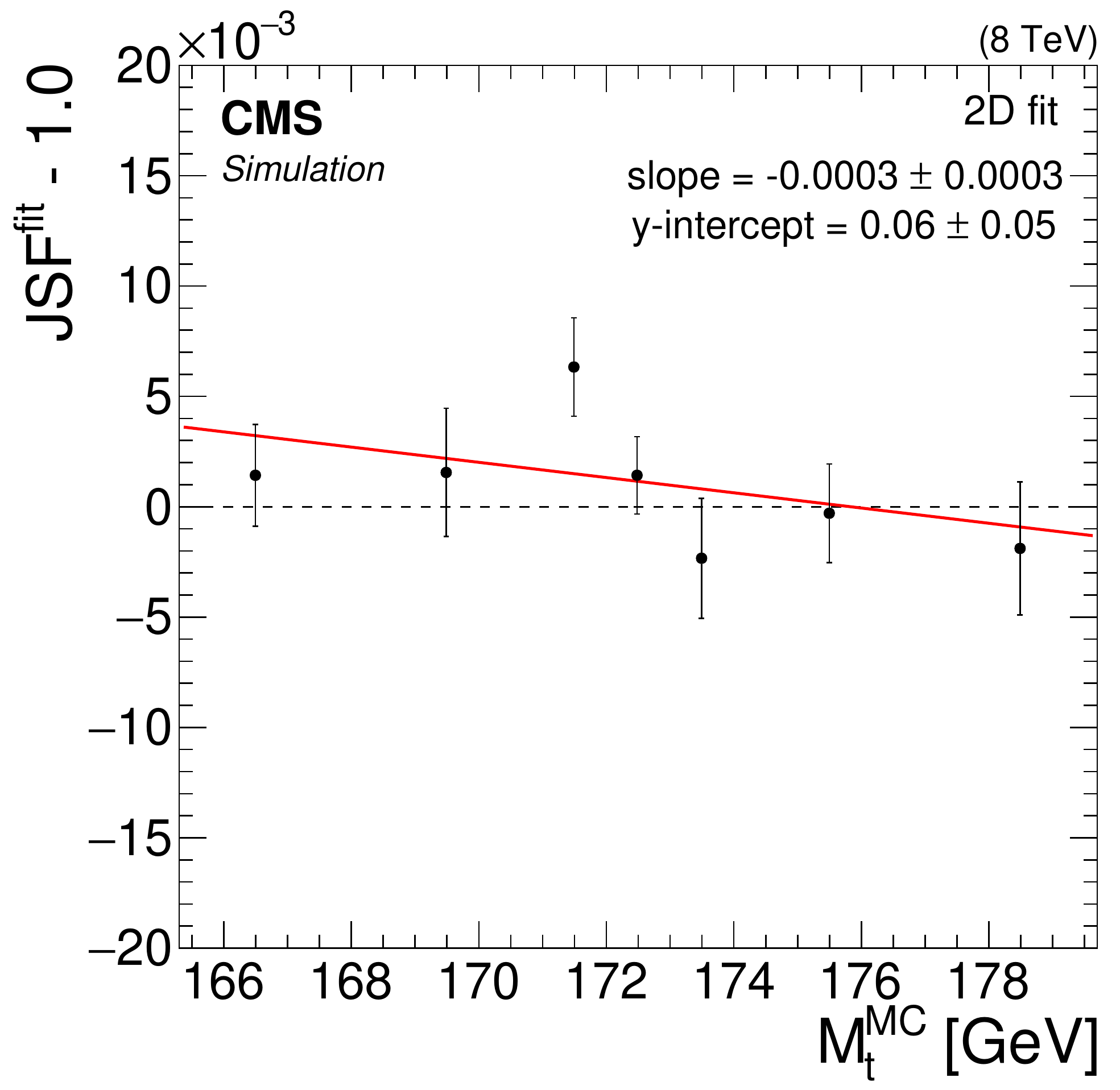} \\
  \includegraphics[width=\cmsFigMultiWidth]{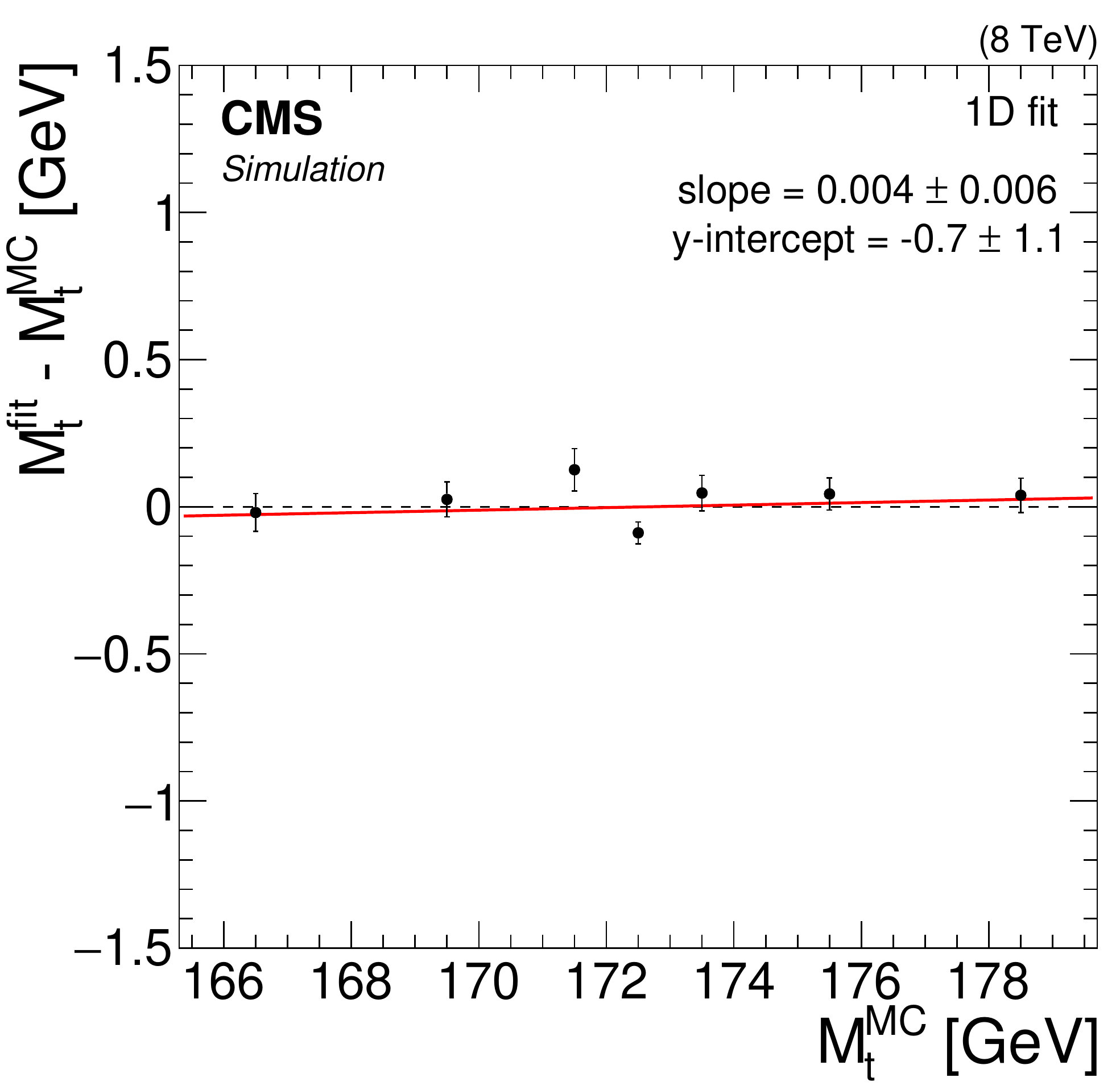}
  \includegraphics[width=\cmsFigMultiWidth]{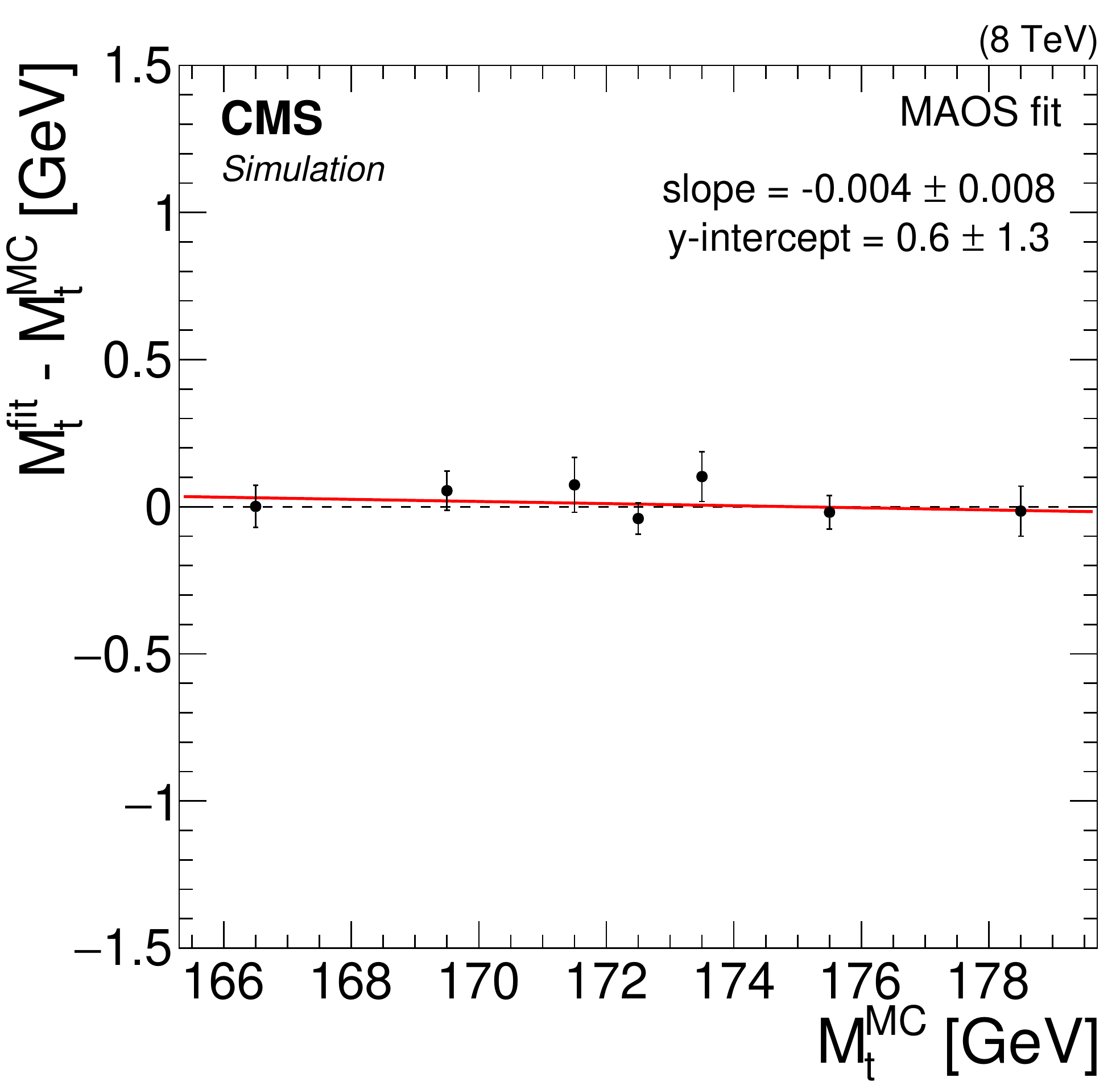} \\
  \includegraphics[width=\cmsFigMultiWidth]{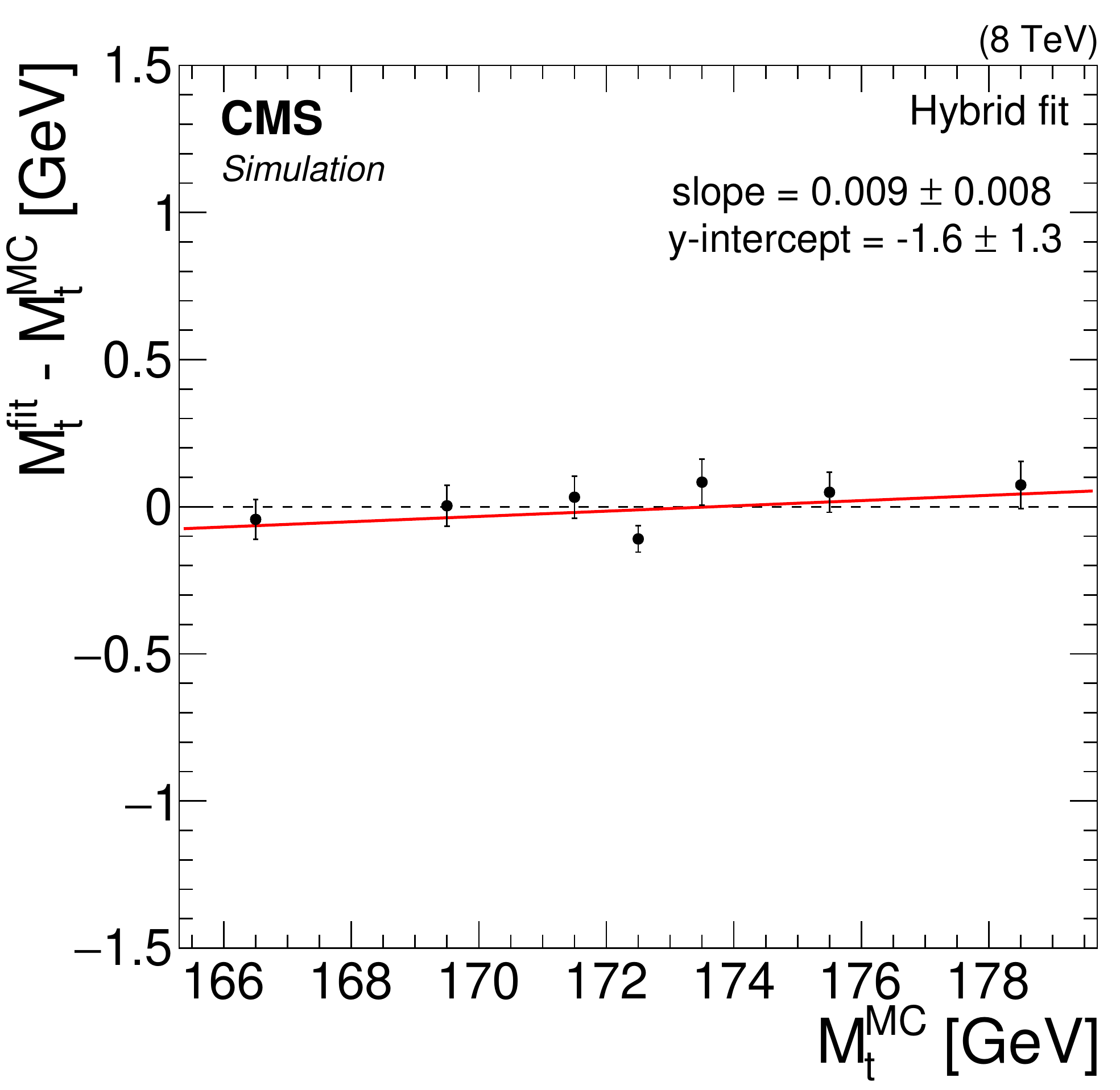}
  \caption{Likelihood fit results as a function of \mtmc\ corresponding to the (top) 2D, (center left) 1D, (center right) MAOS, and (bottom) hybrid fits.  For each value of \mtmc, the fit is conducted using $50$ \PEs\ in MC simulation.  The mean parameter values, $\mt^{\rm fit}$ and $\jsf^{\rm fit}$, are represented by the points, with statistical uncertainties indicated by the error bars.  A best-fit line of the form $y=ax + b$ is shown for each fit configuration.}
  \label{fig:mc_calibration}
\end{figure*}

The performance of the likelihood fitting approach described above is evaluated using events in simulation, where the true values of \mt  and JSF are known.  The fit is conducted using seven different values of \mtmc\ ranging from $166.5$ to $178.5$\GeV for each version of the likelihood fit.  The results of this performance study are shown in Fig.~\ref{fig:mc_calibration}.  The likelihood fits are consistent with zero bias, showing that the GP shape modeling technique accurately captures the distribution shapes and their evolution over several values of \mtmc.  For this reason, no calibration of the fit is necessary for an unbiased determination of the \mt  and JSF parameters.

\subsection*{Combination of 1D and 2D fits}
\label{subsec:comb}

The 1D and 2D fits discussed above have differing sensitivities to various sources of systematic uncertainty in this measurement.  Although the 2D fit successfully mitigates the JES uncertainties, which dominate in the 1D fit, other uncertainties in the 2D method are larger and cause the total precision to worsen (Section~\ref{sec:systematics}).
The best overall precision on the value of \mt  is provided by a \textbf{hybrid fit}, defined as a linear combination of the 1D and 2D fits.  The measured value of \mt  in the hybrid fit is given by:
\begin{linenomath*}
  \begin{equation}
    \label{eq:whyb}
    \mthyb = \whyb\mtD + (1-\whyb)\mtDD,
  \end{equation}
\end{linenomath*}
where the parameter \whyb\ determines the relative weight between the 1D and 2D fits in the combination.  The value of \mthyb\ and its statistical uncertainty are extracted using bootstrap \PEs, as described above.  In each \PE, the measured value of \mthyb\ is given by the linear combination in Eq.~\eqref{eq:whyb} of the measured \mtD\ and \mtDD\ values.
A value of $\whyb = 0.8$ is found to achieve the best precision on \mt  when both statistical and systematic uncertainties are taken into account.  The performance of the hybrid fit, evaluated using MC samples corresponding to seven values of \mtmc, is shown in Fig.~\ref{fig:mc_calibration}.

\section{Systematic uncertainties}
\label{sec:systematics}

The systematic uncertainties evaluated in this measurement are given in Table~\ref{tab:systematics}.  The uncertainties include experimental effects from detector calibration and object reconstruction, and modeling effects mostly arising from the simulation of QCD processes.  All uncertainties are determined by conducting the likelihood fit using events from MC simulation with the relevant parameters varied by ${\pm}1\Delta$, where $\Delta$ is the uncertainty on a particular parameter.  The difference in the measured top quark mass ($\delta\mt$) or JSF ($\delta\jsf$) is taken to be the corresponding systematic uncertainty.  For uncertainties that are evaluated by comparing two or more independent MC samples, the values of $\delta\mt$ and $\delta\jsf$ may be subject to statistical fluctuations.  For this reason, if the value of $\delta\mt$ or $\delta\jsf$ is smaller than its statistical uncertainty in a particular systematic variation, the statistical uncertainty is quoted as the systematic uncertainty.  Finally, if a systematic uncertainty is one-sided, where both $+\Delta$ and $-\Delta$ variations produce $\delta\mt$ or $\delta\jsf$ shifts of the same sign, the larger shift is taken as the symmetric systematic uncertainty.

In the hybrid fit, the systematic uncertainties are evaluated according to the linear combination in Eq.~\eqref{eq:whyb}.
For each systematic variation, this gives $\delta\mthyb = \whyb\delta\mtD + (1-\whyb)\delta\mtDD$.
This approach provides the smallest overall uncertainty, with the largest contributions stemming from the JES, \PQb quark fragmentation modeling, and hard scattering scale.  The next most precise result is given by the 1D fit, also dominated by the same sources of uncertainty.
The JES uncertainties are successfully mitigated in the 2D fit.
The 2D fit, however, is more sensitive to the uncertainties in the top quark \pt spectrum, matching scale, and underlying event tune, so the total systematic uncertainty for the 2D fit is larger than that of the 1D fit.
The MAOS fit has a larger total systematic uncertainty than the 1D fit due to its sensitivity to the JES, top quark \pt spectrum, and \PQb quark fragmentation modeling uncertainties.
Further details on each source of systematic uncertainty are given below.

\begin{table*}[htbp]
  \topcaption{Systematic uncertainties for the 2D, 1D, hybrid, and MAOS likelihood fits.
  The breakdown of JES and \PQb quark fragmentation uncertainties into separate components is shown, where the components are added in quadrature to obtain the total.  The `up' and `down' variations are given separately, with the sign of each variation indicating the direction of the corresponding shift in \mt  or JSF.  The \ca\ character highlights the uncertainty sources that are large in at least one of the likelihood fits.}
  \centering
 \renewcommand{\arraystretch}{1.5}
 \begin{scotch}{llccccc}
      & & $\delta\mtDD$ & $\delta\jsfDD$ & $\delta\mtD$ & $\delta\mthyb$ & $\delta\mtMAOS$ \\
      & & [\GeVns{}] & & [\GeVns{}] & [\GeVns{}] & [\GeVns{}] \\
      \hline
                             JES (total) & \ca & $^{+0.06}_{-0.10}$ & $^{+0.007}_{-0.006}$ & $^{+0.54}_{-0.55}$ & $^{+0.43}_{-0.46}$ & $^{+0.65}_{-0.70}$ \\

      \hspace{5mm}  {-- In situ}      & & {$^{+0.04}_{-0.04}$} & {$^{-0.002}_{+0.003}$} & {$^{-0.22}_{+0.21}$} & {$^{-0.18}_{+0.17}$} & {$^{-0.28}_{+0.24}$} \\
      \hspace{5mm}  {-- Intercalibration} & & {$^{-0.01}_{+0.01}$} & {$^{<0.001}_{<0.001}$} & {$^{-0.04}_{+0.03}$} & {$^{-0.03}_{+0.03}$} & {$^{-0.04}_{+0.04}$} \\
      \hspace{5mm}  {-- Uncorrelated}     & & {$^{+0.04}_{-0.04}$} & {$^{-0.005}_{+0.005}$} & {$^{-0.39}_{+0.39}$} & {$^{-0.32}_{+0.31}$} & {$^{-0.47}_{+0.47}$} \\
      \hspace{5mm}  {-- Flavor}           & & {$^{+0.02}_{-0.09}$} & {$^{+0.004}_{-0.003}$} & {$^{+0.31}_{-0.32}$} & {$^{+0.25}_{-0.27}$} & {$^{+0.39}_{-0.43}$} \\[2ex]

                                      \PQb quark frag. (total) & \ca & $^{+0.39}_{-0.39}$ & $^{+0.001}_{-0.001}$ & $^{+0.40}_{-0.40}$ & $^{+0.40}_{-0.40}$ & $^{+0.67}_{-0.67}$ \\
          \hspace{5mm} {-- Frag. function} & & {$^{+0.38}_{-0.38}$} & {$^{<0.001}_{<0.001}$} & {$^{+0.38}_{-0.38}$} & {$^{+0.38}_{-0.38}$} & {$^{+0.64}_{-0.64}$} \\
      \hspace{5mm} {-- Branching fraction} & & {$^{+0.07}_{-0.07}$} & {$^{+0.001}_{-0.001}$} & {$^{+0.13}_{-0.13}$} & {$^{+0.12}_{-0.12}$} & {$^{+0.20}_{-0.20}$} \\[2ex]

                       JER &     & $^{-0.03}_{+0.08}$ & $^{+0.001}_{-0.002}$ & $^{+0.01}_{-0.05}$ & $^{<0.00}_{-0.03}$ & $^{+0.04}_{-0.04}$ \\
        Unclustered energy &     & $^{+0.10}_{-0.10}$ & $^{+0.001}_{-0.001}$ & $^{-0.02}_{+0.02}$ & $^{-0.04}_{+0.01}$ & $^{-0.11}_{+0.12}$ \\
                   Pileup &     & $^{-0.06}_{+0.04}$ & $^{<0.001}_{<0.001}$ & $^{-0.06}_{+0.05}$ & $^{-0.06}_{+0.05}$ & $^{-0.06}_{+0.05}$ \\
     Electron energy scale &     & $^{-0.38}_{+0.39}$ & $^{+0.002}_{-0.003}$ & $^{-0.21}_{+0.21}$ & $^{-0.24}_{+0.24}$ & $^{-0.02}_{+0.05}$ \\
       Muon momentum scale &     & $^{-0.11}_{+0.09}$ & $^{+0.001}_{<0.001}$ & $^{-0.06}_{+0.05}$ & $^{-0.07}_{+0.06}$ & $^{<0.01}_{+0.01}$ \\
           Electron Id/Iso &     & $^{+0.07}_{-0.02}$ & $^{-0.001}_{<0.001}$ & $^{+0.03}_{-0.01}$ & $^{+0.03}_{-0.01}$ & $^{+0.01}_{<0.01}$ \\
               Muon Id/Iso &     & $^{<0.01}_{<0.01}$ & $^{<0.001}_{<0.001}$ & $^{<0.01}_{<0.01}$ & $^{<0.01}_{<0.01}$ & $^{<0.01}_{<0.01}$ \\
                 \PQb tagging &     & $^{+0.03}_{-0.03}$ & $^{<0.001}_{-0.001}$ & $^{-0.01}_{+0.01}$ & $^{<0.01}_{<0.01}$ & $^{<0.01}_{<0.01}$ \\
      Top quark \pt reweighting & \ca & $^{+0.93}_{\ \ \ \NA}$ & $^{-0.007}_{\ \ \ \ \NA}$ & $^{+0.40}_{\ \ \ \NA}$ & $^{+0.51}_{\ \ \ \NA}$ & $^{+0.72}_{\ \ \ \NA}$ \\
      Hard scattering scale & \ca & $^{-0.36}_{+0.20}$ & $^{+0.007}_{-0.003}$ & $^{+0.31}_{-0.49}$ & $^{+0.21}_{-0.47}$ & $^{+0.33}_{-0.08}$ \\
            Matching scale & \ca & $^{-0.86}_{+0.30}$ & $^{-0.004}_{+0.008}$ & $^{-0.25}_{+0.11}$ & $^{-0.37}_{+0.12}$ & $^{+0.12}_{-0.12}$ \\
    Underlying event tunes & \ca & $^{+0.56}_{-0.56}$ & $^{+0.007}_{-0.007}$ & $^{+0.08}_{-0.08}$ & $^{+0.11}_{-0.11}$ & $^{+0.09}_{-0.09}$ \\
        Color reconnection &     & $^{+0.06}_{-0.06}$ & $^{+0.001}_{-0.001}$ & $^{+0.15}_{-0.15}$ & $^{+0.13}_{-0.13}$ & $^{+0.16}_{-0.16}$ \\
              ME Generator &     & $^{+0.18}_{-0.18}$ & $^{-0.004}_{+0.002}$ & $^{-0.19}_{+0.07}$ & $^{-0.13}_{+0.07}$ & $^{+0.11}_{-0.07}$ \\
                      PDFs &     & $^{+0.14}_{-0.14}$ & $^{+0.001}_{-0.001}$ & $^{+0.17}_{-0.16}$ & $^{+0.17}_{-0.15}$ & $^{+0.17}_{-0.16}$ \\
        \hline
                     Total &     & $^{+1.31}_{-1.25}$ & $^{+0.015}_{-0.014}$ & $^{+0.91}_{-0.95}$ & $^{+0.89}_{-0.93}$ & $^{+1.27}_{-1.02}$ \\
    \end{scotch}
  \label{tab:systematics}
\end{table*}

\begin{itemize}

  \item {Jet energy scale:}
The JES uncertainty is evaluated separately for four components, which are then added in quadrature \cite{jes_correlations}.  The `Intercalibration' uncertainty arises from the modeling of radiation in the \pt- and $\eta$-dependent JES determination.  The `In situ' category includes uncertainties stemming from the determination of the absolute JES using $\gamma$/Z+jet events.
The `Uncorrelated' uncertainty includes uncertainties due to detector effects and pileup.  Finally, the `Flavor' uncertainty stems from differences in the energy response between different jet flavors -- it is a linear sum of contributions from the light quark, charm quark, bottom quark, and gluon responses, which are estimated by comparing the Lund string fragmentation in \PYTHIA \cite{pythia} and cluster fragmentation in \HERWIG{}++ \cite{herwig} for each type of jet.  All JES uncertainties are propagated into the reconstructed \ptmiss\ in each event.

\item {b quark fragmentation:}
The \PQb quark fragmentation uncertainty includes two components that are implemented using event weights.  The first component stems from the \PQb quark fragmentation function, which can modeled using the Lund fragmentation model in the \PYTHIA Z2$^*$ tune, or tuned to empirical results from the ALEPH \cite{aleph} and DELPHI \cite{delphi} experiments.  This component is evaluated by comparing the measurement results in MC simulation using these two tunes of the \PQb quark fragmentation function, with the difference symmetrized to obtain the corresponding uncertainty.  The second uncertainty component stems from the B hadron semileptonic branching fraction, which has an impact on the \PQb quark JES due to the production of a neutrino.  The corresponding uncertainty is evaluated by repeating the measurement with branching fraction values of $10.05\%$ and $11.27\%$, which are variations about the nominal value of $10.50\%$ and encompass the range of values measured from B hadron decays and their uncertainties \cite{pdg}.  Both uncertainty components are combined in quadrature to obtain the total uncertainty.

\item {Jet energy resolution:}
The energy resolution of jets is known to be underestimated in MC simulation compared to data.  This effect is corrected with a set of scale factors that are used to smear the jet four-vectors to broaden their resolutions.  The scale factors are determined in bins of $\eta$.  Here, they are varied within their uncertainties, which are typically 2.5--5\%.  The effect of these variations is also propagated into the \ptmiss.

\item {Unclustered energy:}
The unclustered energy in each event comprises the low-\pt hadronic activity that is not clustered into a jet.
Here, the scale of the unclustered energy is varied by $\pm$10\% \cite{met}.

\item {Pileup:}
The uncertainty in the number of pileup interactions in MC simulation stems from the instantaneous luminosity in each bunch crossing and the effective inelastic cross section.  In this analysis, the number of pileup interactions in MC is reweighted to match the data.  The pileup uncertainty is evaluated by varying the effective inelastic cross section by $\pm$5\%.

\item {Lepton energy scale:}
The electron energy scale is varied up and down by $0.6\%$ in the ECAL barrel ($\abs{\eta} < 1.48$) and by $1.5\%$ in the ECAL endcap ($1.48 < \abs{\eta} < 3.0$) \cite{electron}.  The muon momentum scale is varied up and down by $0.2\%$.  All variations are propagated into the \ptmiss.

\item {Lepton identification and isolation:}
Event weights are applied to adjust the electron and muon yields in MC simulation to account for differences in the identification and isolation efficiencies between data and simulation.  For muons, the uncertainty is taken to be $0.5\%$ of the identification event weight, and $0.2\%$ of the isolation event weight \cite{muon}.  For electrons, the uncertainties are estimated in bins of \pt and $\eta$, and are approximately 0.1--0.5\% of the combined event weight for identification and isolation \cite{electron}.

\item {\PQb tagging efficiency:}
Event weights are applied to adjust the \PQb jet yields in MC simulation to account for the difference in the \PQb tagging efficiency between data and MC simulation \cite{btagging}.  The uncertainties are evaluated in bins of \pt and $\eta$.

\item {Top quark {\pt} reweighting:}
Event weights are applied in order to compensate for a difference in the top quark \pt spectrum between data and MC simulation \cite{top_pt}.  The uncertainty is evaluated by comparing the measurement in MC simulation with and without the weights applied.  The event weights are not applied in the nominal result.  This uncertainty is one-sided by construction, and is not symmetrized.

\item {Hard scattering scale:}
The factorization scale, $\mu_\mathrm{F}$, determines the threshold separating the parton-parton hard scattering from softer interactions embodied in the PDFs.  The renormalization scale, $\mu_\mathrm{R}$, sets the energy scale at which matrix-element calculations are evaluated.  Both of these scales are set to $\mu_\mathrm{F} = \mu_\mathrm{R} = Q$ in the matrix-element calculation and the initial-state parton shower of the \MADGRAPH samples, where $Q^2 = \mt^2 + \sum\pt^2$.  Here, the sum runs over all additional final state partons in the matrix element.  The values of $\mu_\mathrm{F}$ and $\mu_\mathrm{R}$ are varied simultaneously up and down by a factor of two to estimate the corresponding uncertainty.

\item {Matching scale:}
The matrix element-parton shower matching threshold is used to interface the matrix elements generated in \MADGRAPH with parton showers simulated in \PYTHIA.  Its reference value of 20\GeV is varied up and down by a factor of two.

\item {Underlying event tunes and color reconnection:}
The underlying event tunes affect the modeling of soft hadronic activity that results from beam remnants and multi-parton interactions in each event.  The measurement is conducted with a \ttbar\ sample from MC simulation using the `Perugia 2011' tune.  It is compared to results using samples with the `Perugia 2011 mpiHi' and `Perugia 2011 Tevatron' tunes \cite{perugia} in \PYTHIA, corresponding to an increased and decreased underlying event activity, respectively.  The largest difference is symmetrized to obtain the final uncertainty.  The color reconnection (CR) uncertainty is evaluated by comparing measurement results using \ttbar\ samples with the `Perugia 2011' and `Perugia 2011 no CR' tunes \cite{perugia}, where CR effects are not included in the latter.  The difference is symmetrized to obtain the final uncertainty.

\item {Matrix-element generator:}
The measurement is repeated using MC samples produced with the \POWHEG event generator, which provides a next-to-leading-order calculation of the \ttbar\ production.  These measurement results are compared with the reference \ttbar\ MC sample, generated using \MADGRAPH, to determine the corresponding uncertainty.

\item {Parton distribution functions:}
Initial-state partons are described by PDFs.  The corresponding uncertainty is evaluated by applying event weights in the MC simulation to reflect the CT10 PDF set \cite{pdfs} with 50 error eigenvectors.  The total PDF uncertainty is determined by adding the variations corresponding to these error sets in quadrature.

\end{itemize}

\section{Results and discussion}
\label{sec:results}

\begin{figure*}
  \centering
  \includegraphics[width=\cmsFigMultiWidth]{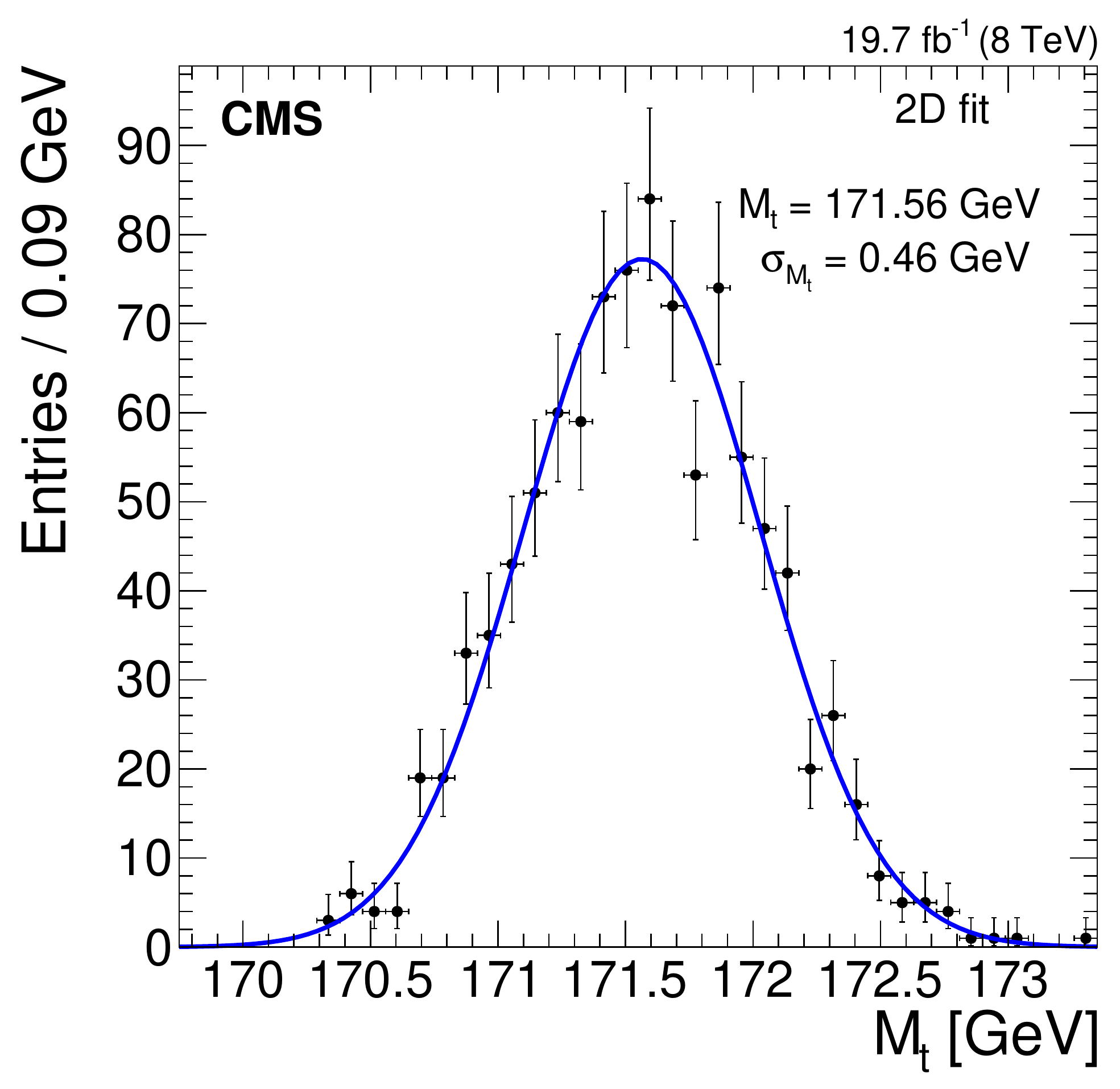}
  \includegraphics[width=\cmsFigMultiWidth]{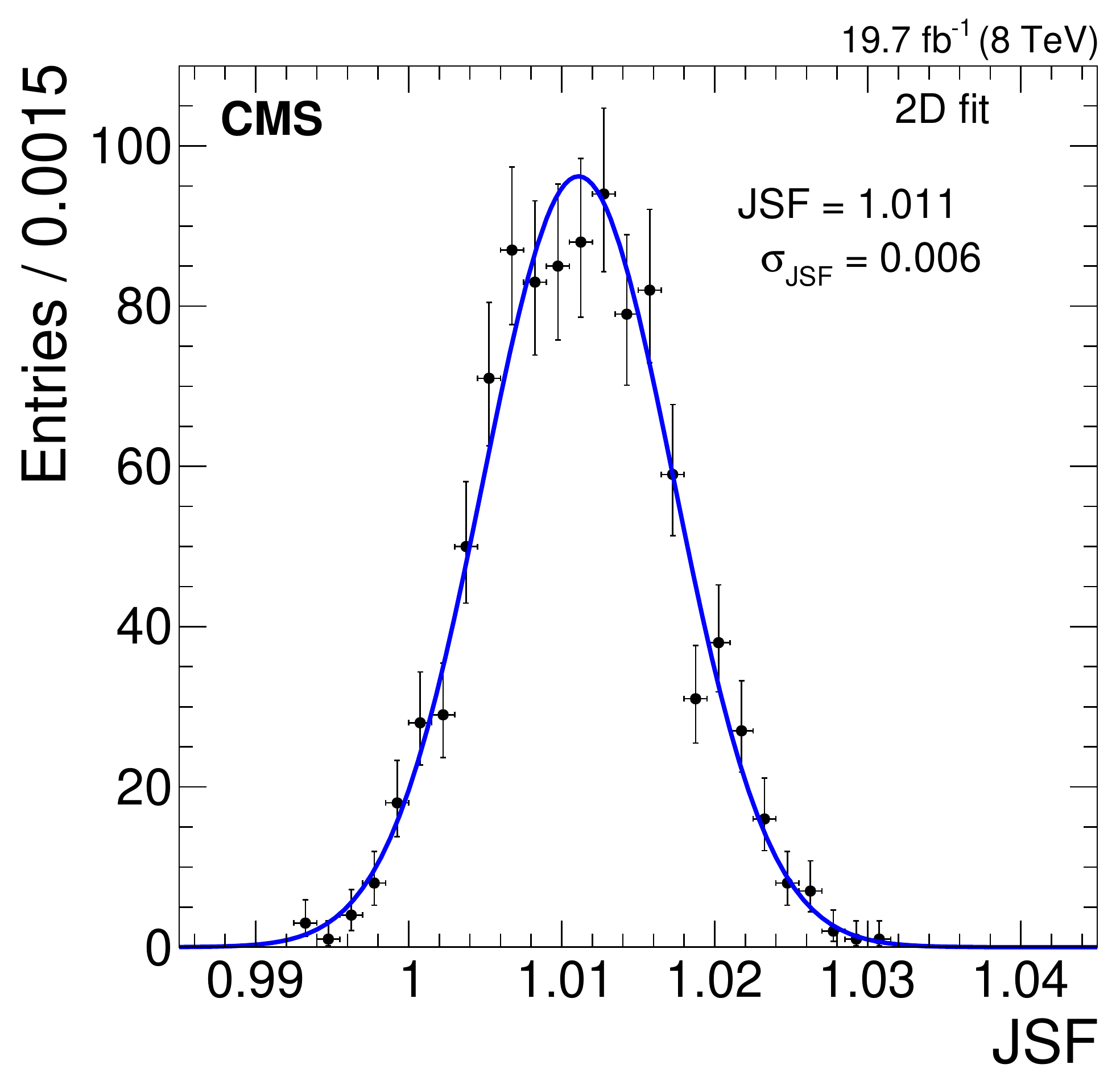} \\
  \includegraphics[width=\cmsFigMultiWidth]{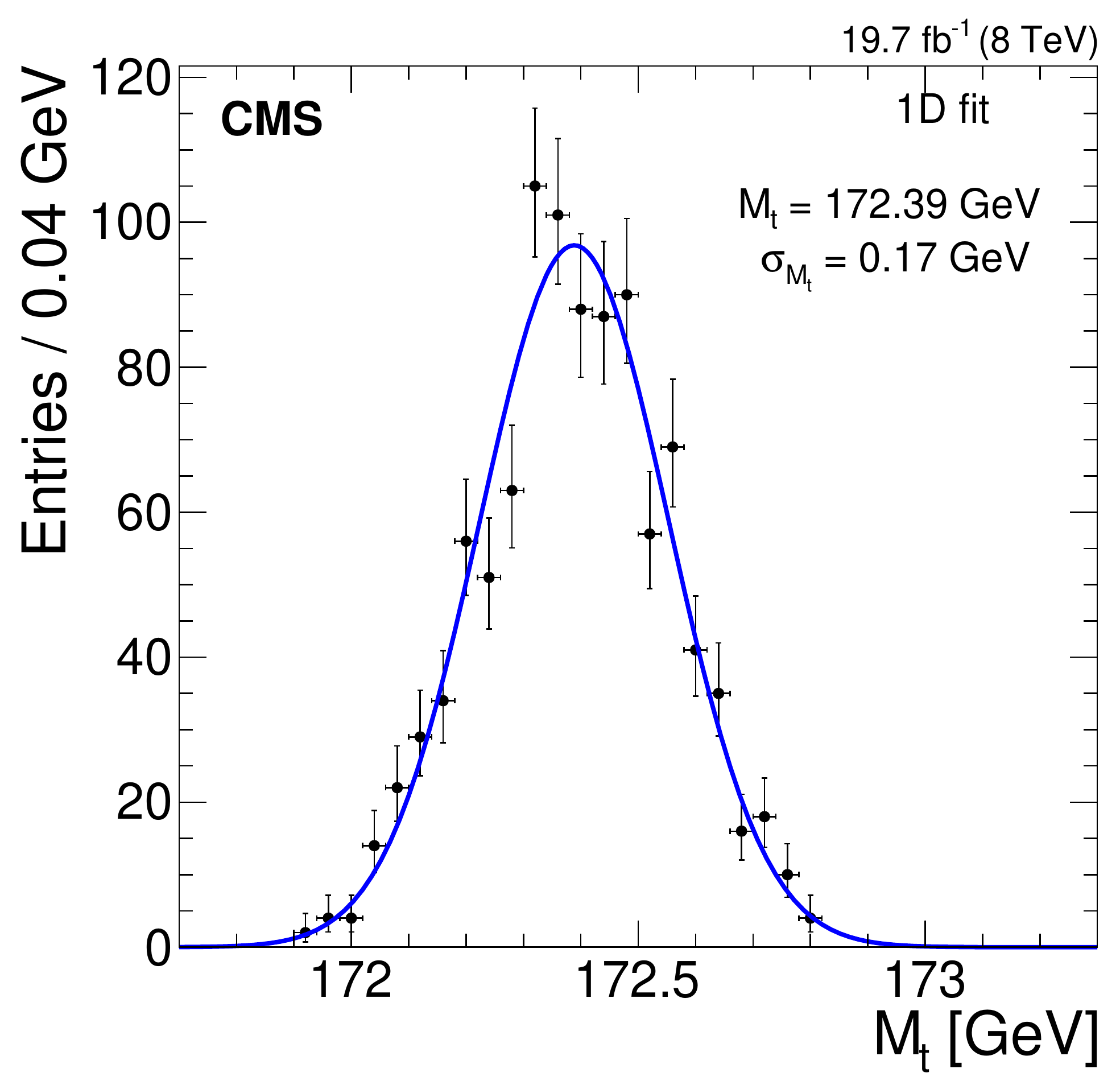}
  \includegraphics[width=\cmsFigMultiWidth]{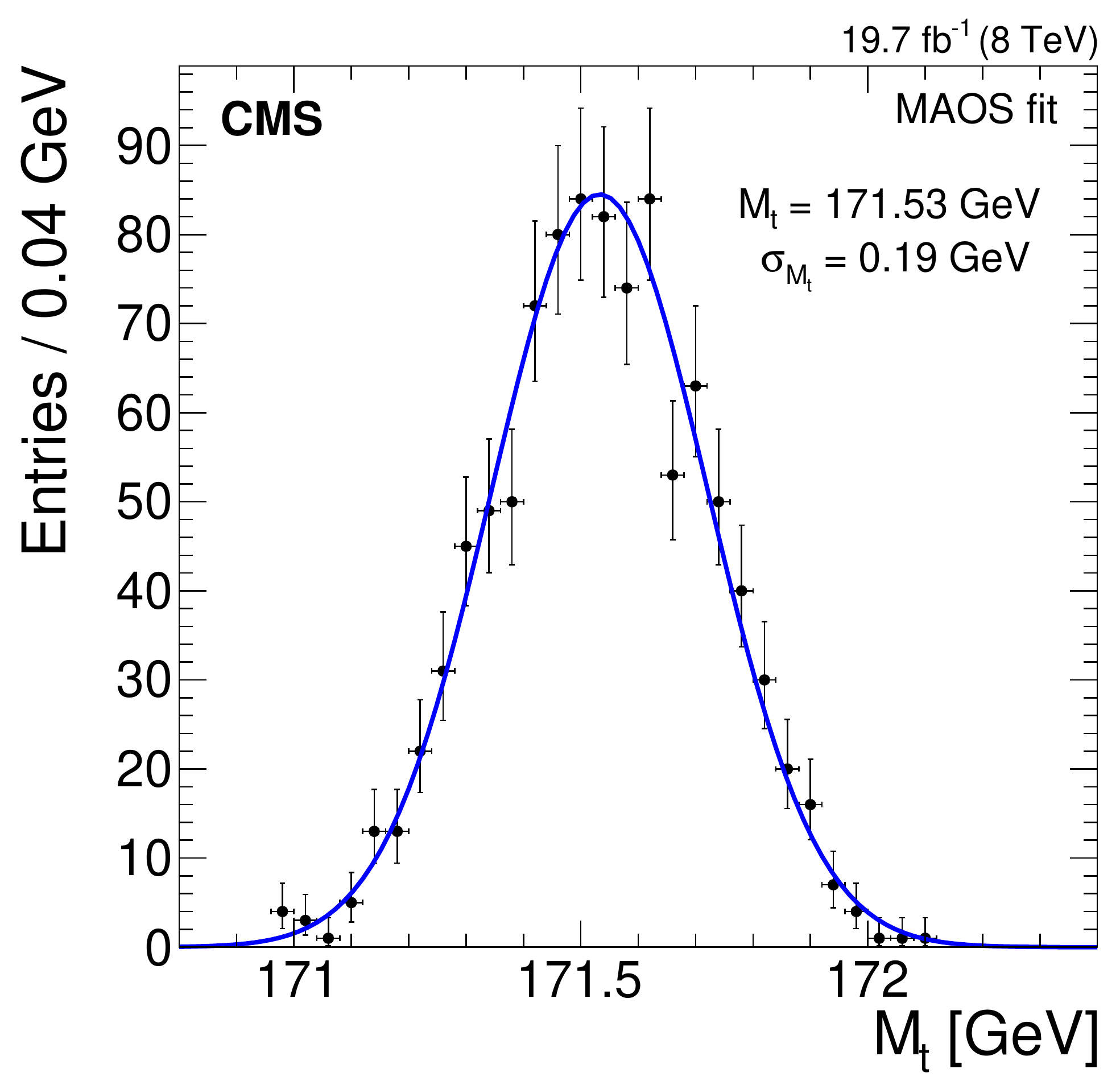} \\
  \includegraphics[width=\cmsFigMultiWidth]{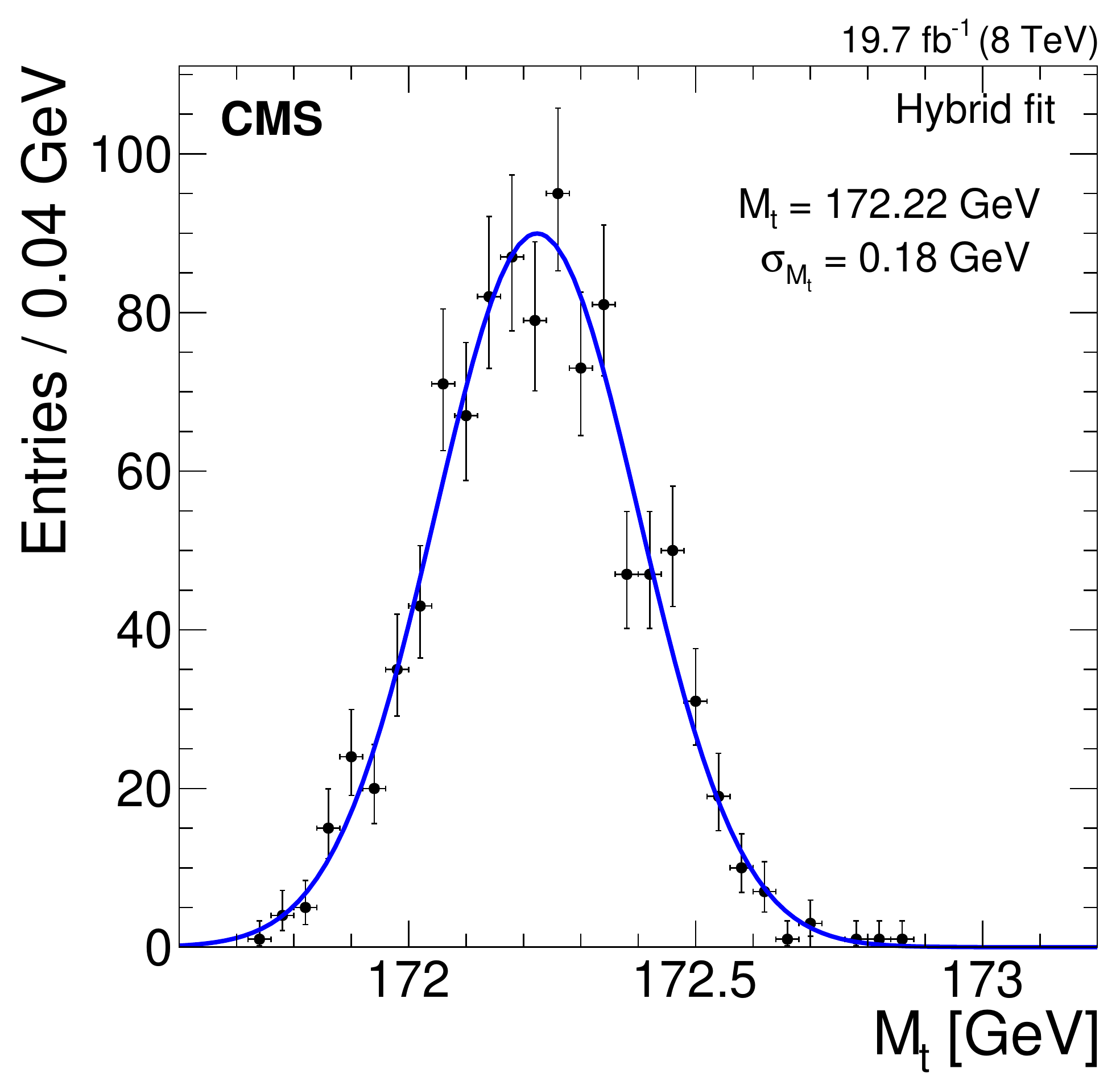}
  \caption{Likelihood fit results using $1000$ bootstrap \PEs\ for the (top) 2D fit, (center left) 1D fit, and (center right) MAOS fit.  (Bottom) hybrid fit results given by the linear combination in Eq.~\eqref{eq:whyb} of the 1D and 2D fits.  The error bars represent the statistical uncertainty corresponding to the number of \PEs\ in each bin. }
  \label{fig:boot_mt}
\end{figure*}

\begin{figure}
  \centering
  \includegraphics[width=0.49\textwidth]{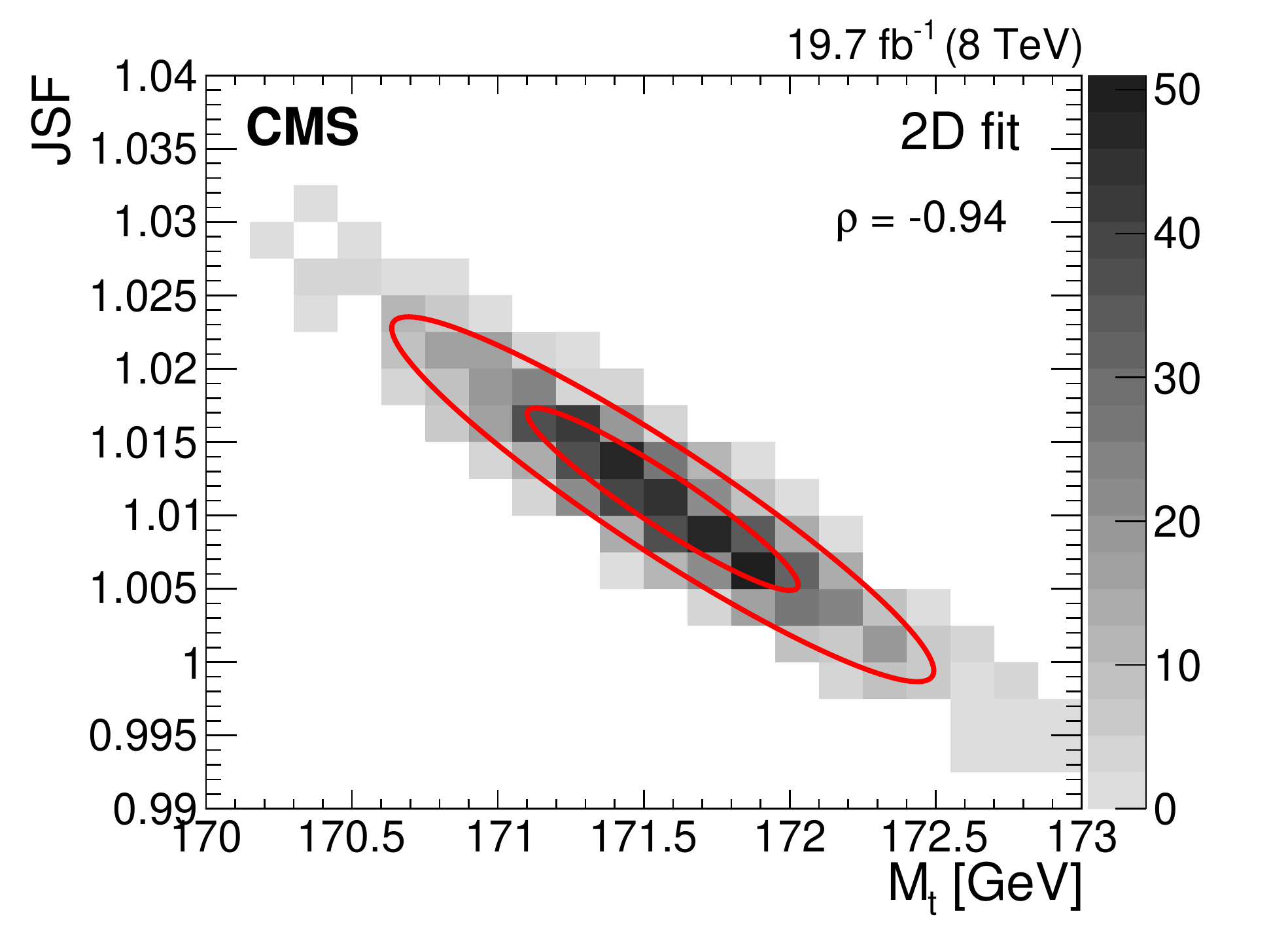}
  \includegraphics[width=0.49\textwidth]{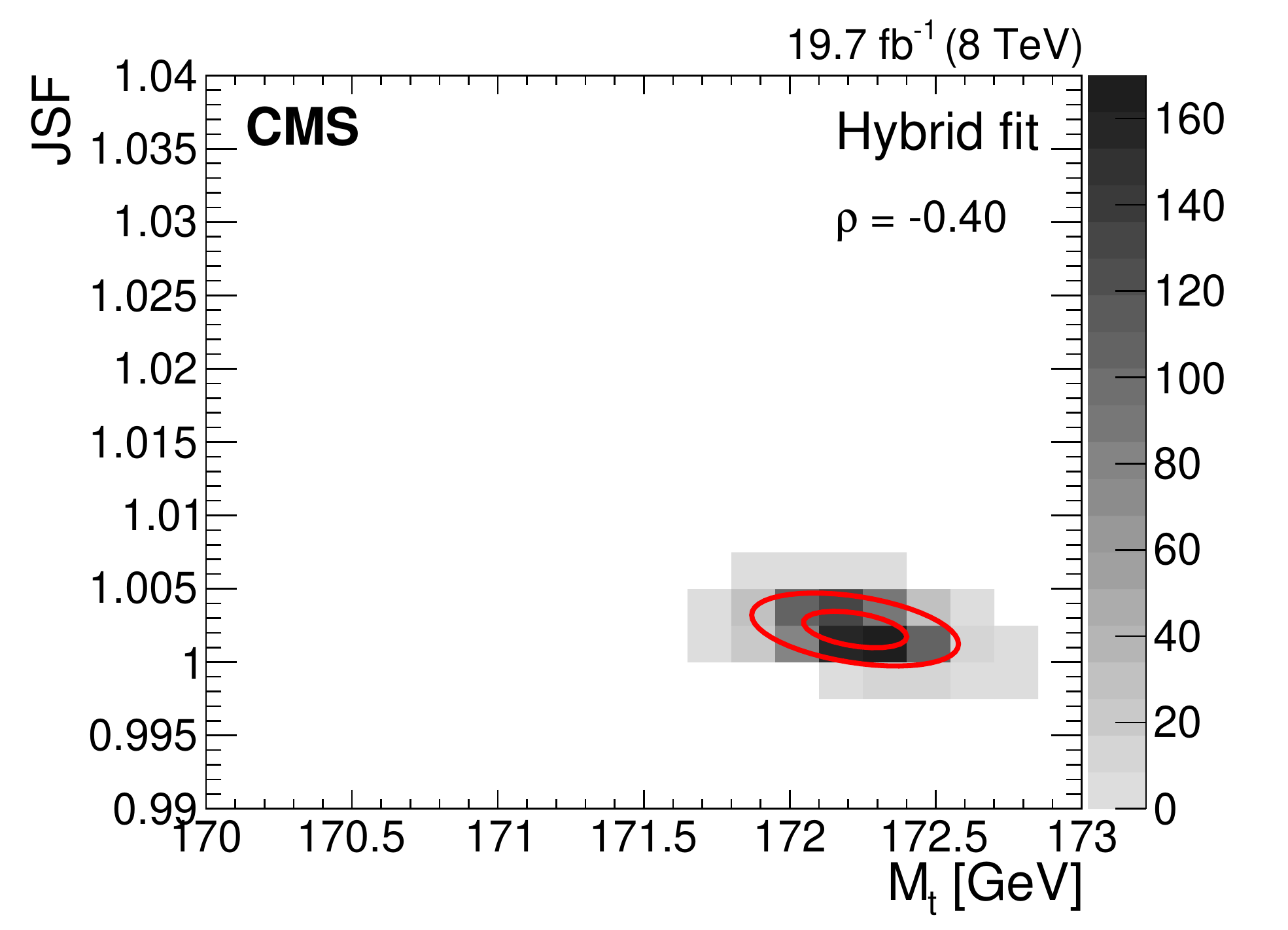}
  \caption{Likelihood fit results corresponding to the 2D fit (\cmsLeft) and hybrid fit (\cmsRight), obtained using $1000$ \PEs\ constructed with the bootstrapping technique.  The shaded histogram represents the number of \PEs\ in each bin of \mt  and \jsf.  Two-dimensional contours corresponding to $-2\Delta\log(\mathcal{L})=1 (4)$ are shown, allowing the construction of one (two) $\sigma$ statistical intervals in \mt  and JSF.  The hybrid fit results are given by a linear combination of the 1D and 2D fit results using Eq.~\eqref{eq:whyb}.}
  \label{fig:mt_vs_jsf}
\end{figure}

\begin{figure}
  \centering
  \includegraphics[width=0.49\textwidth]{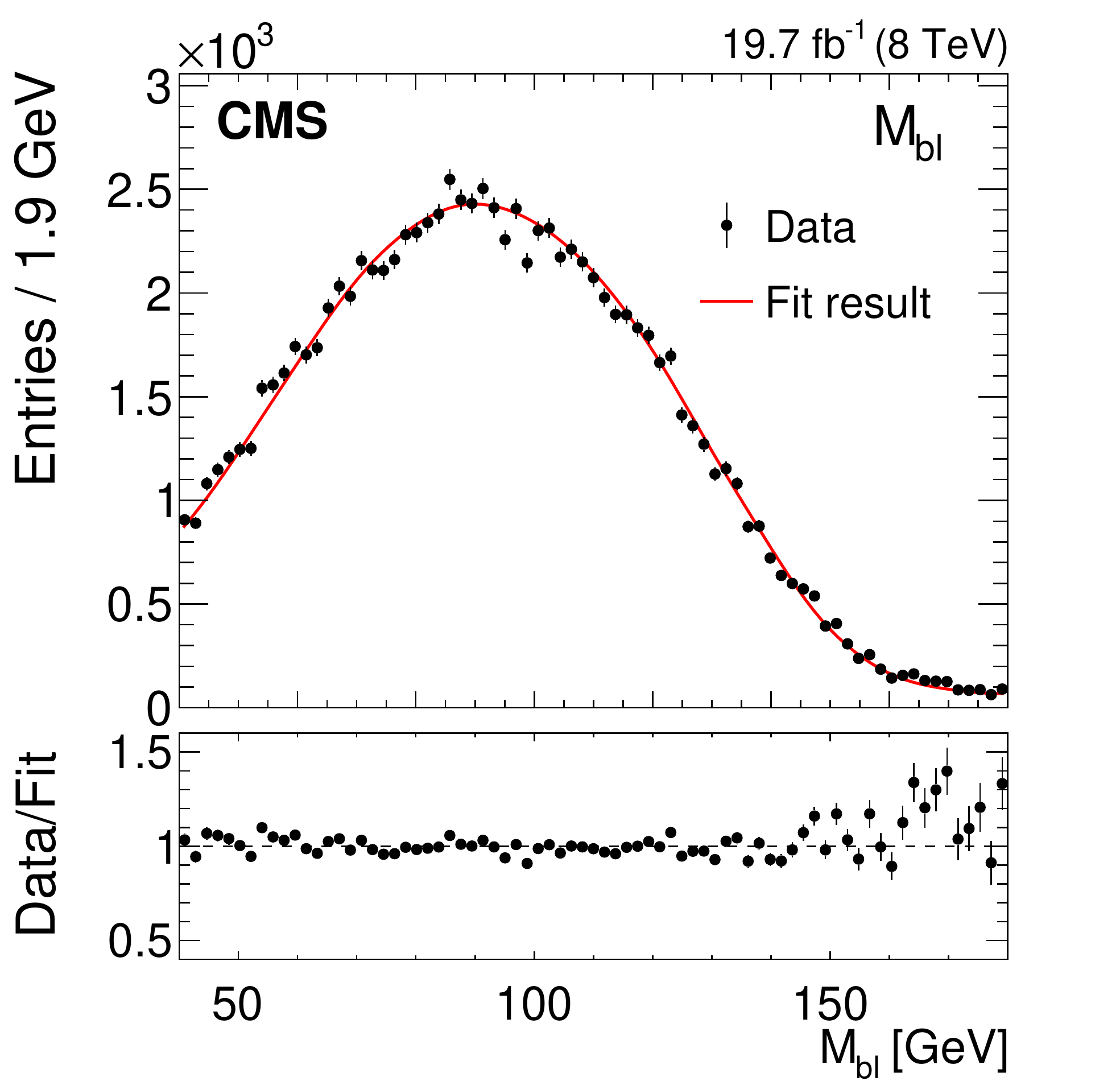}
  \includegraphics[width=0.49\textwidth]{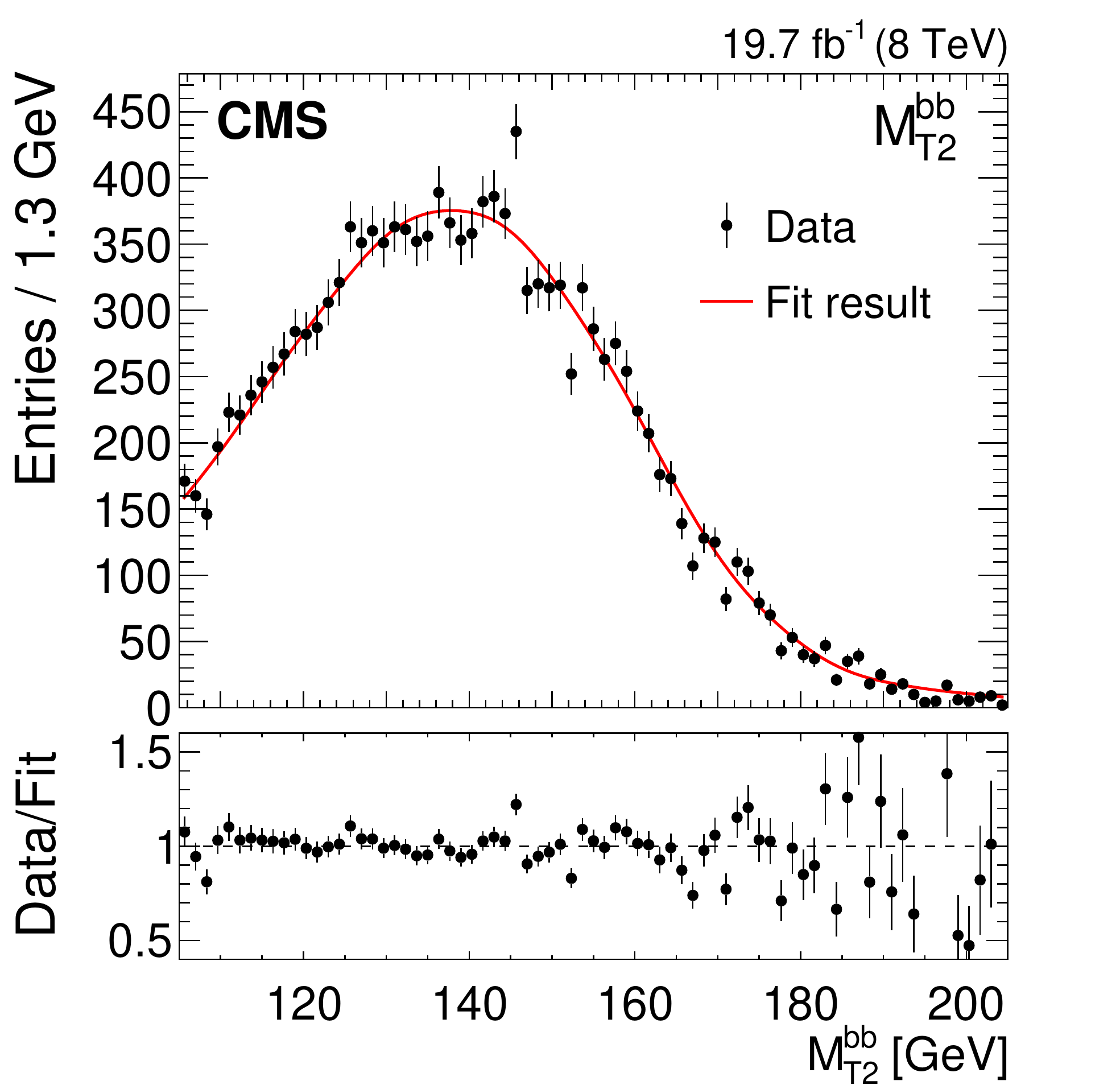}
  \caption{Maximum-likelihood fit result in a typical \PE\ of the 2D likelihood fit in data.  The best fit parameter values for this \PE\ are $\mt=171.99$\GeV and $\jsf = 1.007$.  When the \jsf\ parameter is constrained to be unity in the 1D likelihood fit, the best fit value of \mt  is $172.48\GeV$.  The lower panel shows the ratio between the distribution in data and the best fit distribution in simulation.}
  \label{fig:fitplots_2d}
\end{figure}

{\tolerance=1200
The results for each version of the likelihood fit, determined from 1000 bootstrap \PEs\ in each fit, are shown in Fig.~\ref{fig:boot_mt}.
The 2D fit uses the \mbl\ and \mttbb\ observables to simultaneously determine the values of \mt  and JSF, yielding $\mtDD = \fitDDmtshort$\GeV\ and $\jsfDD = \fitDDjsf$.
The correlation between the \mt  and JSF fit parameters in the 2D fit is shown in Fig.~\ref{fig:mt_vs_jsf}, with a correlation coefficient of $\rho = -0.94$.  The \mbl\ and \mttbb\ distribution shapes corresponding to the fit results in a typical \PE\ are shown in Fig.~\ref{fig:fitplots_2d}.
The 2D fit is successful in mitigating the uncertainty due to the determination of JES, which is otherwise the largest source of systematic uncertainty in this measurement.  In particular, this approach is insensitive to the flavor-dependent component of JES uncertainties --- stemming from differences in the response between \PQb jets, light-quark jets, and gluon jets --- since predominantly \PQb jets are used for the determination of both \mt  and JSF parameters.  The underlying strategy, rooted in a simultaneous fit of two distributions with differing sensitivities to the JSF, does not rely on any specific assumptions about the event topology or final state.  For this reason, it can be a viable option for JES uncertainty mitigation in a variety of physics scenarios.
\par}

The 1D fit is also based on the \mbl\ and \mttbb\ observables, but constrains the JSF parameter to unity.  The 1D fit gives a value of $\mtD = \fitDmt$.
In this approach, the JES accounts for the largest source of uncertainty.  However, other uncertainties are reduced with respect to the 2D fit, resulting in an improved overall precision.

The best overall precision is given by the hybrid fit, which is given by a linear combination of the 1D and 2D fit results.  The 1D and 2D fits use the same set of events and an identical likelihood function constructed from the \mbl\ and \mttbb\ observables.  These fits are fully correlated, with the only difference between them stemming from the treatment of the JSF parameter, which is fixed to unity in the 1D fit and acts as a free parameter in the 2D fit.  The choice to fix the JSF parameter or allow it to float has an impact on the fit sensitivity to a variety of uncertainty sources in addition to the JES.  A linear combination of the 1D and 2D fits with $\whyb = 0.8$, as defined in Eq.~\eqref{eq:whyb}, achieves an optimal balance between all uncertainty sources, thus providing the best overall precision.  The hybrid fit gives:
\begin{linenomath*}
  \begin{align*}
    \mthyb = \fithybmt.
  \end{align*}
\end{linenomath*}
The correlation between the \mt  and JSF fit parameters in the hybrid fit is shown in Fig.~\ref{fig:mt_vs_jsf}, with a correlation coefficient of $\rho = -0.40$.

\begin{figure}
  \centering
  \includegraphics[width=0.49\textwidth]{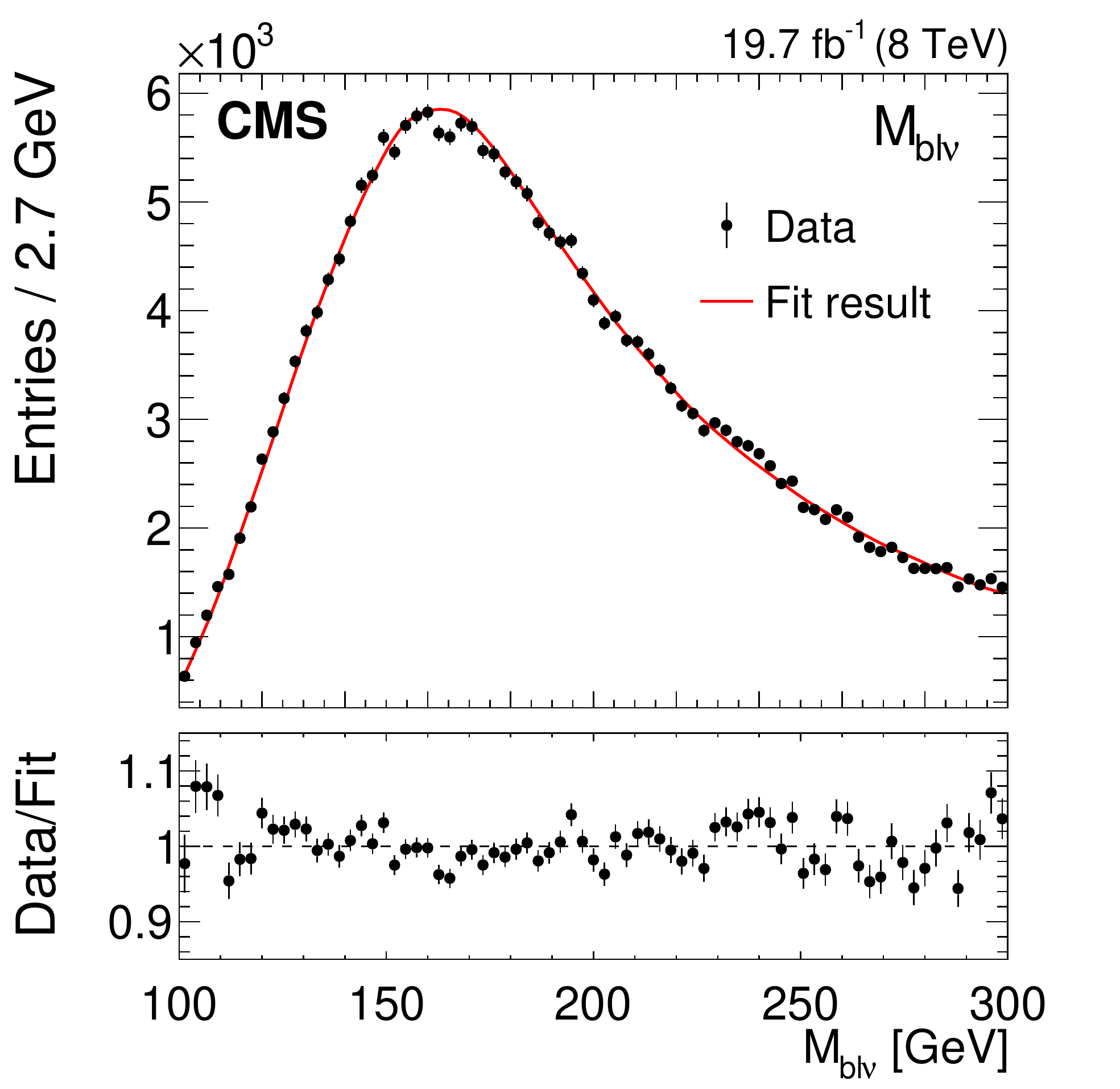}
  \caption{The MAOS \mblv\ distribution corresponding to the maximum-likelihood fit result in a typical \PE\ of the MAOS likelihood fit in data.  The best fit value of \mt  for this \PE\ is $171.54$\GeV.  The lower panel shows the ratio between the distribution in data and the best fit distribution in simulation.}
  \label{fig:fitplots_maos}
\end{figure}

\begin{figure}[htbp]
  \centering
  \includegraphics[width=\cmsFigWidth]{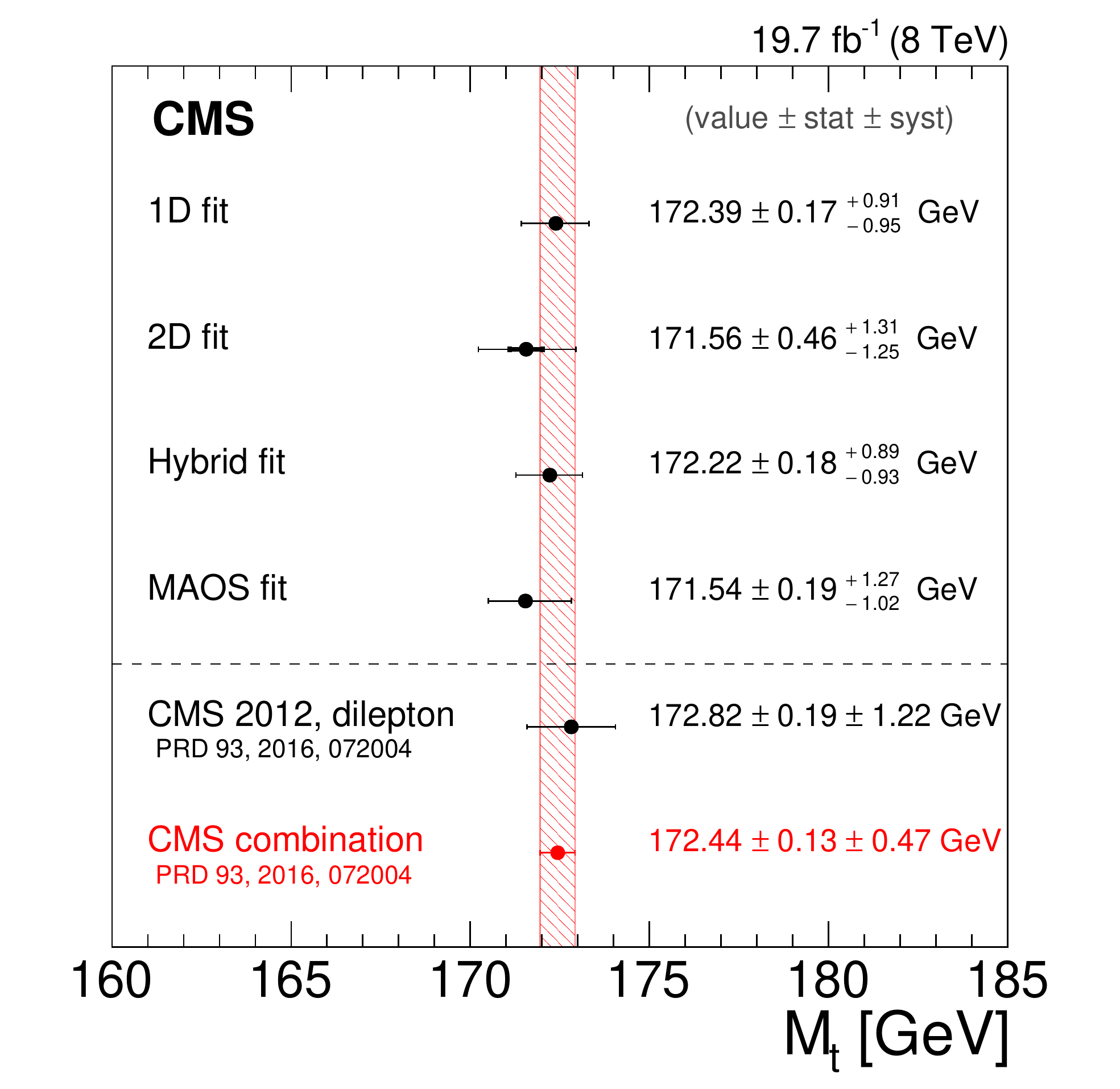}
  \caption{Summary of the 1D, 2D, hybrid, and MAOS likelihood fit results using the 2012 data set at $\sqrt{s}=8\TeV$, corresponding to an integrated luminosity of \luminosity.  A recent dileptonic channel measurement using the 2012 dataset and the most recent combination of \mt  measurements by CMS in all \ttbar decay channels \cite{cms_comb} are shown below the dashed line for reference.}
  \label{fig:mtop_results}
\end{figure}

The MAOS fit substitutes the \mbl\ observable for an \mblv\ invariant mass, yielding a value of $\mtMAOS = \fitMAOSmt$.
The MAOS observable presents a new approach for mass reconstruction in a decay topology characterized by underconstrained kinematics.  Here, the MAOS fit provides a determination of \mt  that is complementary to the 2D, 1D, and hybrid fits.
The MAOS \mblv\ distribution shape corresponding to the fit results in a typical \PE\ is shown in Fig.~\ref{fig:fitplots_maos}.
The results for each version of the likelihood fit are summarized in Fig.~\ref{fig:mtop_results}.

\section{Summary}
\label{section:conclusions}

A measurement of the top quark mass (\mt) in the dileptonic \ttbar\ decay channel is performed using proton-proton collisions at $\sqrt{s}=8\TeV$, corresponding to an integrated luminosity of \luminosity.  The measurement is based on the mass observables \mbl, \mttbb, and \mblv, which allow for mass reconstruction in decay topologies that are kinematically underconstrained.  The sensitivity of these observables to the value of \mt  is investigated using a Fisher information density technique.  The observables are employed in three versions of an unbinned likelihood fit, where a Gaussian process technique is used to model the corresponding distribution shapes and their evolution in \mt  and an overall jet energy scale factor (JSF).  The Gaussian process shapes are nonparametric, and allow for a likelihood fitting framework that gives unbiased results.
The 2D fit provides the first simultaneous measurement of \mt  and \jsf\ in the dileptonic channel.  It is robust against uncertainties due to the determination of jet energy scale, including the flavor-dependent uncertainty component arising from differences in the response between \PQb\ jets, light-quark jets, and gluon jets.  The fit yields $\mt = \fitDDmt$ and $\jsf = \fitDDjsf$.  The most precise measurement of \mt  is given by a linear combination of this result with a fit in which the JSF is constrained to be unity, yielding a value of \fithybmt.
This measurement achieves a 25\% improvement in overall precision on \mt  compared to previous dileptonic channel analyses using the 2012 data set at CMS.  The improvement can be attributed to a reduction of the systematic uncertainties in the measurement.

\begin{acknowledgments}
We congratulate our colleagues in the CERN accelerator departments for the excellent performance of the LHC and thank the technical and administrative staffs at CERN and at other CMS institutes for their contributions to the success of the CMS effort. In addition, we gratefully acknowledge the computing centers and personnel of the Worldwide LHC Computing Grid for delivering so effectively the computing infrastructure essential to our analyses. Finally, we acknowledge the enduring support for the construction and operation of the LHC and the CMS detector provided by the following funding agencies: BMWFW and FWF (Austria); FNRS and FWO (Belgium); CNPq, CAPES, FAPERJ, and FAPESP (Brazil); MES (Bulgaria); CERN; CAS, MoST, and NSFC (China); COLCIENCIAS (Colombia); MSES and CSF (Croatia); RPF (Cyprus); SENESCYT (Ecuador); MoER, ERC IUT, and ERDF (Estonia); Academy of Finland, MEC, and HIP (Finland); CEA and CNRS/IN2P3 (France); BMBF, DFG, and HGF (Germany); GSRT (Greece); OTKA and NIH (Hungary); DAE and DST (India); IPM (Iran); SFI (Ireland); INFN (Italy); MSIP and NRF (Republic of Korea); LAS (Lithuania); MOE and UM (Malaysia); BUAP, CINVESTAV, CONACYT, LNS, SEP, and UASLP-FAI (Mexico); MBIE (New Zealand); PAEC (Pakistan); MSHE and NSC (Poland); FCT (Portugal); JINR (Dubna); MON, RosAtom, RAS, RFBR and RAEP (Russia); MESTD (Serbia); SEIDI, CPAN, PCTI and FEDER (Spain); Swiss Funding Agencies (Switzerland); MST (Taipei); ThEPCenter, IPST, STAR, and NSTDA (Thailand); TUBITAK and TAEK (Turkey); NASU and SFFR (Ukraine); STFC (United Kingdom); DOE and NSF (USA).
If acknowledgements for individuals are required for a short letter because some of the principal authors are funded through individual grants, it should be OK to add the lines below concerning individuals even for a short letter, but please first consult with the PubComm chair.

\hyphenation{Rachada-pisek} Individuals have received support from the Marie-Curie program and the European Research Council and EPLANET (European Union); the Leventis Foundation; the A. P. Sloan Foundation; the Alexander von Humboldt Foundation; the Belgian Federal Science Policy Office; the Fonds pour la Formation \`a la Recherche dans l'Industrie et dans l'Agriculture (FRIA-Belgium); the Agentschap voor Innovatie door Wetenschap en Technologie (IWT-Belgium); the Ministry of Education, Youth and Sports (MEYS) of the Czech Republic; the Council of Science and Industrial Research, India; the HOMING PLUS program of the Foundation for Polish Science, cofinanced from European Union, Regional Development Fund, the Mobility Plus program of the Ministry of Science and Higher Education, the National Science Center (Poland), contracts Harmonia 2014/14/M/ST2/00428, Opus 2014/13/B/ST2/02543, 2014/15/B/ST2/03998, and 2015/19/B/ST2/02861, Sonata-bis 2012/07/E/ST2/01406; the National Priorities Research Program by Qatar National Research Fund; the Programa Clar\'in-COFUND del Principado de Asturias; the Thalis and Aristeia programs cofinanced by EU-ESF and the Greek NSRF; the Rachadapisek Sompot Fund for Postdoctoral Fellowship, Chulalongkorn University and the Chulalongkorn Academic into Its 2nd Century Project Advancement Project (Thailand); and the Welch Foundation, contract C-1845.
\end{acknowledgments}
\bibliography{auto_generated}

\clearpage
\appendix

\section{Statistical sensitivity of kinematic observables}
\label{sec:sensitivity}

The sensitivity of a kinematic observable to the value of a parameter such as \mt  can be quantified by its Fisher information \cite{cramer,dasgupta}.  The Fisher information of an observable is related to its likelihood function, \Lt, which we have introduced in Eq.~\eqref{eq:L} and reproduce here:
\begin{linenomath*}
  \begin{equation}
    \label{eq:Lsens}
    \log \Lt(m) = \sum_i^N{\log\fa(x_i | m)},
  \end{equation}
\end{linenomath*}
where $\fa(x|m)$ is the distribution of observable $x$ normalized to unity over its range, $m$ is a free parameter, and $N$ is the number of observations of $x$.  In this measurement, we have $x =$ \mbl, \mttbb, or \mblv, $m =$ \mt  or \jsf, and $N$ is a multiple of the total number of events.
For simplicity we consider the distribution shape \fa as a function of only one free parameter.
The Fisher information corresponding to the shape $\fa(x|m)$ is given by:
\begin{linenomath*}
  \begin{equation}
    \label{eq:fisher}
    \mathcal{I}(m) = \int{\left(\frac{\partial}{\partial m}\log \fa(x|m)\right)^2 \fa(x|m) \,\rd x}.
  \end{equation}
\end{linenomath*}
The quantity $\mathcal{I}(m)$ provides a measure of curvature near the likelihood maximum.
It can be interpreted as the variance of the slope, $\left(\partial\log\fa(x|m)/\partial m\right)$, known as the `statistical score' of $\fa(x|m)$.

The Fisher information is related to the precision of a measurement by the Cr\'{a}mer-Rao bound:
\begin{linenomath*}
  \begin{equation}
    \label{eq:cramer_rao}
    \sm^2 \geq \frac{1}{N\,\mathcal{I}(m)},
  \end{equation}
\end{linenomath*}
where \sm\ is the statistical uncertainty on parameter $m$.  In a likelihood with large $N$, the shape of the likelihood near its maximum is roughly Gaussian, and the bound approaches an equality.
This expression confirms the expected relationship $\sm \propto 1/\sqrt{N}$ between the statistical uncertainty and the value of $N$, but also reveals the proportionality factor as the reciprocal of the Fisher information.  It expresses the uncertainty \sm\ in terms of the total number of events, the shape \fa, and the derivative $\partial \fa/\partial m$.

The Fisher information also provides a mathematical framework for quantifying the sensitivity of an observable at a specific point on its shape.
In this analysis, the \mbl\ and \mttbb\ observables have kinematic endpoints at approximately $\sqrt{\mt^2-\mw^2}$ and \mt, respectively; the MAOS \mblv\ observable is an invariant mass whose shape contains a peak near the value of \mt.  Because these features carry a dependence on the value of \mt, the regions near the endpoints of \mbl\ and \mttbb\ and the peak of \mblv\ are expected to contribute significantly to the sensitivity of these observables.
To relate these local features to the Fisher information, we consider the integral in Eq.~\eqref{eq:fisher} over the value of observable $x$.
Here, the integrand of the Fisher information can be interpreted as the contribution to the total sensitivity stemming from a specific value of $x$.  Rewriting the integrand in a more convenient form, we define the `local shape sensitivity' function by:
\begin{linenomath*}
  \begin{equation}
    \label{eq:sensitivity}
    s(x|m) \equiv \frac{1}{\fa(x|m)}\left[\frac{\partial \fa(x|m)}{\partial m}\right]^2.
  \end{equation}
\end{linenomath*}
This function is also known as the Fisher information density.  It is shown for the \mbl, \mttbb, and MAOS \mblv\ observables in Figs.~\ref{fig:dist_mbl}, \ref{fig:dist_mt2}, and \ref{fig:dist_maos}, respectively, with $m=\mt$ and the JSF parameter fixed to unity.  It is observed to peak near the kinematic endpoints of \mbl\ and \mttbb, and on the left-side edge of \mblv.
The values of $x$ where $s(x|m) = 0$ coincide with the stationary points at which the distribution shapes in Figs.~\ref{fig:dist_mbl}, \ref{fig:dist_mt2}, and \ref{fig:dist_maos} intersect.  This is a reflection of the fact that in a likelihood fit, events with a value of $x$ near a stationary point make little or no contribution to the determination of $m$.
In general, the shape of $s(x|m)$ for each observable establishes a link between the underlying kinematic properties of the observable and regions of high and low sensitivity on its shape.  In this analysis, it provides heuristic information about the \mbl, \mttbb, and MAOS \mblv\ distributions, and their sensitivity to the value of \mt.

In addition to providing heuristic information, the local shape sensitivity function is used in this analysis to identify potential overfitting effects in the Gaussian process (GP) shapes.  Overfitting occurs when the interpolation between GP training points is not smooth, causing fluctuations in the shape that may be difficult to identify by eye.  Such fluctuations can be a source of bias, both in the determination of \mt  and its corresponding uncertainties.  A typical symptom of overfitting is an under-estimated statistical uncertainty on the value of \mt.  This can occur when fluctuations in the GP shape increase the value of the slope $\partial\fa(x|\mt)/\partial \mt$ appearing in Eq.~\eqref{eq:fisher}, thus artificially increasing the Fisher information of the corresponding shape.  The issue is easily revealed by the shape of $s(x|\mt)$, which acquires visible fluctuations when overfitting is indeed present.  In such cases, overfitting can be mitigated by increasing relevant GP hyperparameter values to improve the smoothness of the GP shape.

\section{Gaussian process regression technique}
\label{sec:gp_appendix}

The likelihood fit described in Section~\ref{sec:fit} uses distribution shapes of the form $\fa(x|\mt,\jsf)$, where $x$ is the value of an observable (\mbl, \mttbb, or \mblv), and \mt  and \jsf\ are free parameters in the fit.
In this analysis, the distribution shapes \fa are modeled with a Gaussian process (GP) regression technique.
We define a point, $\gp_i$, on each distribution shape by its position in $x$, \mt, and \jsf:
\begin{linenomath*}
  \begin{equation}
    \gp_i \equiv (x_i, \mt{}_i, \jsf_i).
  \end{equation}
\end{linenomath*}
The value of the shape at $\gp_i$ is given by $\fa(\gp_i) = \fa(x_i | {\mt}_i, {\jsf}_i)$.  The point $\gp_i$ can be a training point, at which the value of \fa is known and used as an input into the GP regression process; or it can be a test point, at which the value of \fa is to be determined.
Each GP shape is trained using binned distributions of the observable $x$ in MC simulation.  For each observable, $35$ binned distributions are used, corresponding to seven values of \mtmc\ ranging from $166.5$ to $178.5$\GeV and five values of \jsf\ ranging from $0.97$ to $1.03$.  Each distribution has $75$ bins in $x$, yielding a total of $2625$ training points.  This binning scheme is chosen to provide an accurate modeling of the distribution shapes, while mitigating the effects of statistical fluctuations.
The GP regression technique interpolates between the discrete values of $x$, \mt, and \jsf\ covered by these training points to provide a shape that is smooth over its range.

The `Gaussian' in GP refers to the distribution of possible values of the shape \fa.  The value at a single point, $\fa(\gp_i)$, is distributed according to a one-dimensional Gaussian function rather than being treated as an exact quantity.  The mean of this Gaussian function is the most probable value of the shape at that point $\gp_i$, and it is the value used for likelihood fitting (Section~\ref{sec:fit}); the variance stems from the modeling uncertainty inherent in the GP regression process.  The values $\fa(\gp_i)$ and $\fa(\gp_j)$ at any two points follow a two-dimensional Gaussian distribution and are related by a covariance.  The correlation between $\fa(\gp_i)$ and $\fa(\gp_j)$ determines the degree to which the GP shape is allowed to vary between the points $\gp_i$ and $\gp_j$.  By extension, any $N$ values of the shape are described by an $N$-dimensional Gaussian distribution, and are related by an $N\times N$ covariance matrix.  To determine the value of the shape at a test point $\gp_{N+1}$, an $(N+1)$-dimensional Gaussian distribution is constructed relating the training point values $\fa(\gp_1)\ldots \fa(\gp_N)$ to the test point value $\fa(\gp_{N+1})$.  Then, $\fa(\gp_1)\ldots \fa(\gp_N)$ are fixed to their known values, and the $(N+1)$-dimensional Gaussian distribution is reduced to a one-dimensional conditional Gaussian distribution representing the possible values of $\fa(\gp_{N+1})$.

\begin{figure}
  \centering
  \includegraphics[width=0.5\textwidth]{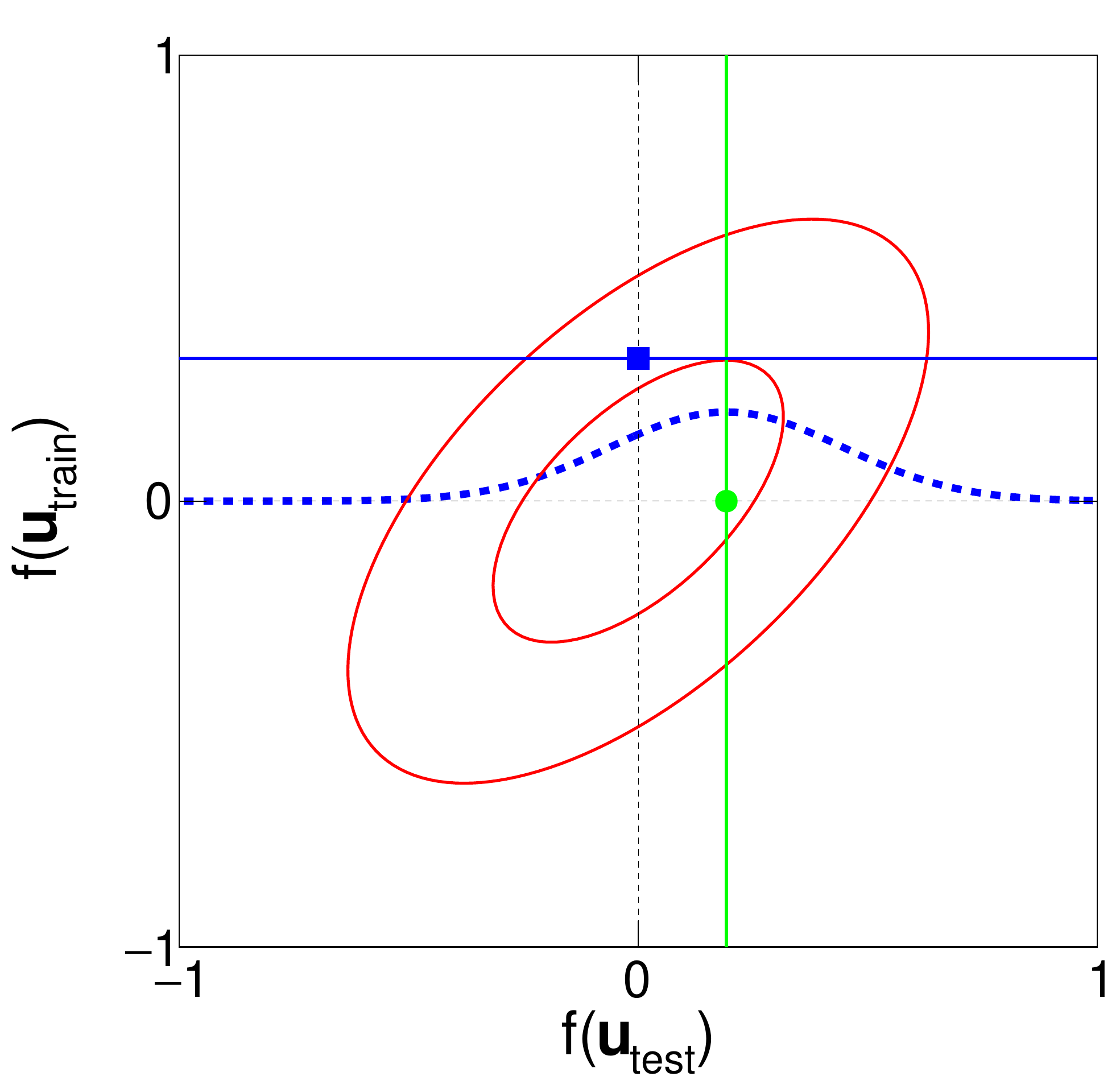}
  \caption{Demonstration of the GP conditioning process, given in Eqs.~\eqref{eq:gp_mean} and \eqref{eq:gp_sigma}, for one training point and one test point.  The covariance between the value of the shape at the training and test point is represented by the ellipse.  The known value of the shape at the training point (square point) determines the mean value of the shape at the test point (round point and vertical line).  The distribution of possible values of the shape at the test point is represented by the dashed curve.}
  \label{fig:gp_conditioning}
\end{figure}

To demonstrate this process graphically, we consider a simple GP with one training point, \gptrain, at which the value $\fa(\gptrain)$ is known, and one test point, \gptest, at which the value $\fa(\gptest)$ is to be evaluated.  The values of $\fa(\gptrain)$ and $\fa(\gptest)$ follow a two-dimensional Gaussian prior distribution with mean values $\mu_{\text{train}}$ and $\mu_{\text{test}}$, and a covariance represented by:
\begin{linenomath*}
  \begin{equation}
    \label{eq:gp_covariance}
    \mathbf{C} = \left[
      \begin{array}{cc}
        \sigma_{\text{train}}^2 & \rho\sigma_{\text{train}}\sigma_{\text{test}} \\
        \rho\sigma_{\text{train}}\sigma_{\text{test}} & \sigma_{\text{test}}^2
      \end{array}
    \right],
  \end{equation}
\end{linenomath*}
where $\sigma_{\text{train}}^2$ and $\sigma_{\text{test}}^2$ are the variances of $\fa(\gptrain)$ and $\fa(\gptest)$, and $\rho$ is the correlation coefficient.  We set $\mu_{\text{train}} = \mu_{\text{test}} = 0$ to reflect our zero prior knowledge of \fa over its range.  The resulting joint Gaussian distribution is represented by the contours in Fig.~\ref{fig:gp_conditioning}.
To evaluate the shape \fa at the test point, we fix $\fa(\gptrain)$ to its known value, indicated by the square point in Fig.~\ref{fig:gp_conditioning}.  The possible values of $\fa(\gptest)$ are now constrained to lie along the horizontal line, giving rise to the conditional Gaussian distribution indicated by the dashed curve.
The mean of the conditional Gaussian is taken to be the value of the shape at the test point.

In this analysis, the conditioning process described above is generalized to $N+1$ dimensions to accommodate all $N$ training points and one test point at which the shape \fa is evaluated.
The mean, $\mu_{N+1}$, and variance, $\sigma_{N+1}^2$, of \fa at test point $\gp_{N+1}$ are given by:
\begin{linenomath*}
  \begin{align}
    \label{eq:gp_mean}
    \mu_{N+1} &= \mathbf{k}^T \mathbf{C}_N^{-1} \mathbf{t}, \\
    \label{eq:gp_sigma}
    \sigma_{N+1}^2 &= c - \mathbf{k}^T \mathbf{C}_N^{-1} \mathbf{k},
  \end{align}
\end{linenomath*}
\sloppy where $\mathbf{t}$ is a column vector containing the $\fa(\gp_i)$ values for all $N$ training points, \mbox{$\mathbf{k} = \cov(\fa(\gp_i),\fa(\gp_{N+1}))$} is the covariance between the value of \fa at the $i$th training point and the value at the test point, and $c = \cov(\fa(\gp_{N+1}),\fa(\gp_{N+1}))$.  The matrix $\mathbf{C}_N = \cov(\fa(\gp_i),\fa(\gp_j))$ is the $N\times N$ covariance matrix expressing the joint Gaussian distribution between the values of \fa at all $N$ training points.
In this analysis, the value of \fa at each point is given by the mean defined in Eq.~\eqref{eq:gp_mean}.  The variance in Eq.~\eqref{eq:gp_sigma} is provided here for completeness.

The covariance $\cov(\fa(\gp_i),\fa(\gp_j))$ between any two points is determined by a kernel function that is set by the analyzer.  The kernel function defines the covariance matrix $\mathbf{C}_N$ in Eqs.~\eqref{eq:gp_mean} and \eqref{eq:gp_sigma}, and its properties determine the smoothness characteristics of the final shape.  A conventional choice for the GP kernel function is a Gaussian---this ensures that the correlation between any two points is suppressed at a large separation.  In practice, the kernel is a three-dimensional function that controls the smoothness of the shape along $x$, \mt, and JSF.  It also includes a correlation term between \mt  and JSF to reflect the kinematic relationship between them.  The result is a product of a one-dimensional Gaussian (controlling the smoothness along $x$) with a two-dimensional Gaussian (controlling the smoothness along \mt  and JSF).  For any two points $\gp_i$ and $\gp_j$ on the shape, the kernel is given by:
\begin{linenomath*}
  \begin{multline}
    \label{eq:gp_kernel}
      \cov(\fa(\gp_i),\fa(\gp_j)) = N_1 \left[ N_2 \exp\left\{-\frac{1}{2\theta_1^2}(x_i-x_j)^2\right\} \right. \\
                               \left. \exp\left\{-\frac{1}{2(1-\rho^2)}\left(\frac{1}{\theta_2^2}({\mt}_i-{\mt}_j)^2 + \frac{1}{\theta_3^2}(\jsf_i-\jsf_j)^2\right. \right. \right. \\
                               \left. \left. \left. - \frac{2\rho}{\theta_2\theta_3}({\mt}_i-{\mt}_j)(\jsf_i-\jsf_j)\right)\right\} + \sigma_i^2\delta_{ij}\right].
  \end{multline}
\end{linenomath*}
Here, $N_1$, $N_2$, $\theta_1$, $\theta_2$, $\theta_3$, and $\rho$ are the GP hyperparameters, $\sigma_i$ is a noise parameter that accounts for the statistical uncertainty on the distribution bin underlying each training point, and $\delta_{ij}$ is the Kronecker delta function.  The terms inside the exponentials specify the covariance between any two values of the shape as a function of their corresponding $x$, \mt, and \jsf.  The hyperparameters $\theta_1$, $\theta_2$, and $\theta_3$ specify the length scales over which the GP shape is allowed to vary, and $\rho$ is a correlation coefficient that couples the \mt  and \jsf\ parameters.  The hyperparameter $N_1$ specifies the overall normalization of the kernel function, and $N_2$ determines the relative normalization between the Gaussian and noise terms.

The values of all hyperparameters are determined with the help of a cross-validation likelihood fit \cite{rasmussen+williams}, conducted for each observable separately.  The length scale hyperparameters ($\theta_1$, $\theta_2$, and $\theta_3$) must be small enough for the GP shape to pass through the training points, and large enough for the shape to interpolate smoothly between them. Hyperparameters that are under-estimated satisfy the former criterion, but cause overfitting to occur in the resulting GP shape.  This creates a noisy interpolation between training points, and may lead to bias in the measured value of \mt  and its uncertainties.  In this analysis, the GP shapes are checked for overfitting effects using the local shape sensitivity function described in Appendix~\ref{sec:sensitivity}.

\cleardoublepage \section{The CMS Collaboration \label{app:collab}}\begin{sloppypar}\hyphenpenalty=5000\widowpenalty=500\clubpenalty=5000\textbf{Yerevan Physics Institute,  Yerevan,  Armenia}\\*[0pt]
A.M.~Sirunyan, A.~Tumasyan
\vskip\cmsinstskip
\textbf{Institut f\"{u}r Hochenergiephysik,  Wien,  Austria}\\*[0pt]
W.~Adam, E.~Asilar, T.~Bergauer, J.~Brandstetter, E.~Brondolin, M.~Dragicevic, J.~Er\"{o}, M.~Flechl, M.~Friedl, R.~Fr\"{u}hwirth\cmsAuthorMark{1}, V.M.~Ghete, C.~Hartl, N.~H\"{o}rmann, J.~Hrubec, M.~Jeitler\cmsAuthorMark{1}, A.~K\"{o}nig, I.~Kr\"{a}tschmer, D.~Liko, T.~Matsushita, I.~Mikulec, D.~Rabady, N.~Rad, B.~Rahbaran, H.~Rohringer, J.~Schieck\cmsAuthorMark{1}, J.~Strauss, W.~Waltenberger, C.-E.~Wulz\cmsAuthorMark{1}
% ...
\vskip\cmsinstskip
% ...
1:~~Also at Vienna University of Technology, Vienna, Austria\\
% ...

\end{sloppypar}
\end{document}